\documentclass[%
 reprint,
 amsmath,amssymb,
 aps,
]{revtex4-2}

\usepackage[utf8]{inputenc}
\usepackage{graphicx}
\usepackage{svg}
\svgsetup{inkscapeexe=inkscape, inkscapearea=drawing, inkscapeversion=1}

\usepackage{color}
\usepackage{amssymb}
\usepackage{amsmath}
\usepackage{listings}
\PassOptionsToPackage{hyphens}{url}\usepackage{hyperref}
\usepackage{float}
\usepackage{subfig}
\usepackage{adjustbox}
\usepackage{algorithm}
\usepackage{algpseudocode}
\usepackage{tikz}
\usetikzlibrary{quantikz}
\usepackage{soul}
\usepackage{bm}
\usepackage{braket}
\usepackage[english]{babel}
\usepackage{array, tabularx, multirow, makecell,lipsum,environ}
\usepackage{amsthm}
\usepackage{appendix}
\usepackage{listings}
\usepackage{enumitem}
\usepackage{xcolor}

\DeclareMathOperator*{\argmin}{argmin}

\begin{document}

\preprint{APS/123-QED}

\title{Multi-Objective Optimization and Network Routing with Near-Term Quantum Computers}

\author{Shao-Hen Chiew}
\email{shaohen@entropicalabs.com}
\affiliation{Entropica Labs, 186b Telok Ayer Street, Singapore 068632}

\author{Kilian Poirier}
\affiliation{Entropica Labs, 186b Telok Ayer Street, Singapore 068632}

\author{Rajesh Mishra}
\email{Current affiliation : Department of Physics, University of Illinois at Urbana-Champaign, 1110 West Green Street, Urbana, Illinois 61801-3080, USA}
\affiliation{Entropica Labs, 186b Telok Ayer Street, Singapore 068632}

\author{Ulrike Bornheimer}
\affiliation{Entropica Labs, 186b Telok Ayer Street, Singapore 068632}

\author{Ewan Munro}
\affiliation{Entropica Labs, 186b Telok Ayer Street, Singapore 068632}

\author{Si Han Foon}%
\affiliation{Defence Science and Technology Agency, 1 Depot Road, Singapore 109679}%

\author{Christopher Wanru Chen}%
\affiliation{Defence Science and Technology Agency, 1 Depot Road, Singapore 109679}%

\author{Wei Sheng Lim}%
\affiliation{Defence Science and Technology Agency, 1 Depot Road, Singapore 109679}%

\author{Chee Wei Nga}%
\affiliation{Defence Science and Technology Agency, 1 Depot Road, Singapore 109679}%

\begin{abstract}
Multi-objective optimization is a ubiquitous problem that arises naturally in many scientific and industrial areas. Network routing optimization with multi-objective performance demands falls into this problem class, and finding good quality solutions at large scales is generally challenging. In this work, we develop a scheme with which near-term quantum computers can be applied to solve multi-objective combinatorial optimization problems. We study the application of this scheme to the network routing problem in detail, by first mapping it to the multi-objective shortest path problem. Focusing on an implementation based on the quantum approximate optimization algorithm (QAOA) -- the go-to approach for tackling optimization problems on near-term quantum computers -- we examine the Pareto plot that results from the scheme, and qualitatively analyze its ability to produce Pareto-optimal solutions. We further provide theoretical and numerical scaling analyses of the resource requirements and performance of QAOA, and identify key challenges associated with this approach. Finally, through Amazon Braket we execute small-scale implementations of our scheme on the IonQ Harmony 11-qubit quantum computer.
\end{abstract}

\maketitle

\tableofcontents

\newpage

\section{Introduction}
Multi-objective optimization problems (MOOPs) arise naturally in many scientific and industrial areas, where the interplay between multiple conflicting objectives gives rise to a set of optimal solutions, rather than a unique one. They are especially prevalent in engineering contexts, where complex systems involving multiple objectives are encountered \cite{collette2004multiobjective,boyd2004convex,emmerich2018tutorial}. In particular, we focus on the case where the state space and hence the set of optimal solutions is discrete, where they are referred to as multi-objective combinatorial optimization problems (MCOPs). An example of such a MCOP is the multi-objective network routing problem, which asks for paths between specified source and destination nodes in a graph that are optimal with respect to multiple objectives. This optimization problem arises in contexts such as the design of Wireless Ad-Hoc/Sensor Networks \cite{akyildiz2002survey,pantazis2007survey,sarangapani2017wireless,karl2007protocols} and next-generation communication networks \cite{kulkarni2010computational,bockelmann2018towards,sharma2019toward}, where large-scale networks with multiple requirements have to be simultaneously satisfied to meet performance demands.

However, the solution of such problems is generally difficult, as finding the global optima -- also known as Pareto-optimal solutions -- of general MCOPs is NP-hard \cite{back1996evolutionary,coello2007evolutionary}. Physics- and biology-inspired meta-heuristic classical algorithms have been particularly successful in this area in the past two decades \cite{deb2002fast,dorigo2006ant,kennedy2006swarm} due to their scalable computational costs, and are the go-to methods for industrial use-cases. Even so, due to the importance and complexity of MCOPs, there is a need to develop theoretical and algorithmic tools to solve large-scale problems in more memory and time efficient ways.

Besides developments in classical computers, quantum computers are currently experiencing an explosive growth, both in theory and realization. Noisy intermediate-scale quantum (NISQ) computers with qubit counts in the small hundreds are now available, and there is a growing body of work investigating their ability to solve challenging optimization problems in areas ranging from chemistry \cite{tilly2022variational} to finance \cite{egger2020quantum}, with the hope that they can outperform existing classical algorithms in the near future. In particular, there has been a strong focus on variational quantum algorithms (VQAs) such as the Quantum Approximate Optimization Algorithm (QAOA) \cite{farhiqaoa}, which can be used to tackle combinatorial optimization problems with existing quantum computers. A natural question is whether similar methods can be applied to to efficiently obtain high quality solutions to MCOPs such as the network routing problem, to satisfy the performance demands of next-generation wireless networks.

In this context, a significant challenge is the fact that, for a given use-case, investigating the performance of VQAs such as QAOA typically requires an empirical approach. However, given the relatively small size of currently available quantum computers and the complexity of simulating quantum computations classically, the scope for experimentation is currently rather limited. Nonetheless, in this work we adopt a pragmatic approach, using the standard form of QAOA \cite{farhiqaoa} as a solution method for MCOPs. This permits us to: (a) use small-scale problem instances to draw insights on how the QAOA solution relates to Pareto-optimal solutions of the MCOP, and; (b) analyse the scalability of the QAOA approach against known generic limitations of VQAs.


The key results of our work are summarized as follows:
\begin{enumerate}
    \item We develop a general framework with which near-term quantum computers can be used to solve MCOPs, by producing multiple Pareto-optimal solutions in both \textit{a priori} (with the preferences of the decision maker taken into account prior to optimization) and \textit{a posteriori} (independent of the decision maker's preferences) manners. This is achieved by casting the multiple objectives and constraints of a MCOP in Quadratic Unconstrained Binary Optimization (QUBO) form, and scalarizing it to obtain a cost function which can be variationally optimized with a VQA. For small problem instances, visualization of the output quantum state on Pareto plots provides insights which we analyze and explain in a qualitative manner.
    
    \item We provide a formulation of the network routing problem that is amenable to implementation with QAOA on near-term quantum computers. Using results from graph theory, we determine its resource requirements, and provide a scaling analysis. In particular, we show that this encoding scheme possesses resource requirements that scale mildly with the connectivity of the underlying graph, which is in principle compatible with resources available on current NISQ hardware.

    \item Numerical simulations of small problem instances using standard QAOA (as described in Ref. \cite{farhiqaoa}) show that this framework can produce high-quality solutions efficiently. Concretely, by increasing the circuit depth, we observe a correspondingly proportional increase in the probability of successfully obtaining Pareto-optimal solutions. However, as we explain in Section \ref{subsec:QAOA_mixer_initial_state}, in terms of problem size the efficacy of the QUBO-based QAOA approach is limited by the large fraction of infeasible solutions in the underlying search space, resulting from the presence of optimization constraints.
    
    \item We run a number of demonstrative test cases on the 11-qubit IonQ Harmony quantum computer, accessed through Amazon Braket. The results obtained are in clear agreement with those of numerical simulations.
\end{enumerate}

This article is structured as follows. We begin by fully describing our approach to obtain optimal solutions of a MCOP and discuss interpretations of resulting Pareto plots in Section.~\ref{sec:qc_moop}. We then describe the network routing problem, a concrete example of a generally difficult MCOP which we study throughout the article, along with relevant objectives that arise from reasonable assumptions in Section.~\ref{sec:network_routing}. This is followed by theoretical and numerical analyses on the scaling of resources and performance of the procedure in Section.~\ref{sec:analyses}, and experiments on quantum computers in Section.~\ref{sec:qpu_results}. 

We refer the reader to the appendices for a survey of relevant work in the usage of near-term quantum algorithms in solving MCOPs and the shortest path problem (Appendix.~\ref{appendix:survey}), a review of MOOPs (Appendix.~\ref{appendix:moop}), the mathematical formulation of the multi-objective shortest-path problem (Appendix.~\ref{appendix:shortest}), graph theoretic ideas used in this work (Appendix.~\ref{appendix:graph_theory}), the formulation of the network routing problem along with complete specifications of physical parameters used to model its objectives (Appendix.~\ref{appendix:network}), specifications of problem instances considered in our numerical calculations and experiments (Appendix.~\ref{appendix:problem_instances} and \ref{appendix:fig_problem}), additional numerical results for the chosen parameter initialization scheme (Appendix.~\ref{appendix:initializations}), and additional discussions on the Pareto plot (Appendix.~\ref{appendix:Feasible_Infeasible}).

Numerical simulations and the set-up of quantum computations in this work were performed through OpenQAOA \cite{sharma2022openqaoa}, an open-source Python package tailored for QAOA and its variants. 

\section{Solving MCOPs with Quantum Computers} \label{sec:qc_moop}

In this section, we describe our approach to solve MCOPs using near-term quantum computers with VQAs such as the QAOA, starting with brief introductions to the QAOA, MCOPs, and combinatorial optimization/QUBO problems. For more in-depth discussions on the above topics, we refer the reader to \cite{farhiqaoa}, \cite{lucas2014ising} and \cite{boyd2004convex,emmerich2018tutorial} respectively, along with the Appendices.

A key component of our procedure is the QAOA. Belonging to the class of VQAs, it was introduced to allow NISQ computers to provide approximate solutions to combinatorial optimization problems such as graph partitioning, coloring, and constraint satisfaction problems \cite{farhiqaoa}. A QUBO problem can be defined by an upper triangular matrix with real values. It is specified within QAOA through a corresponding cost Hamiltonian $H_\text{c}$ expressed as a sum of $l$ local terms $H_\text{c} = \sum_{i=1}^{l} h_i$, which can be obtained by converting the binary variables of the QUBO problem to Ising variables \cite{lucas2014ising}. This allows the quantum state $\ket{\psi(\vec{\beta}, \vec{\gamma})}$ to be prepared by a $p$-layer state-preparation ansatz $U_{\text{QAOA}}(\vec{\beta}, \vec{\gamma})$:
\begin{align} \label{eq:qaoa_unitary}
\begin{split}
  \ket{\psi(\vec{\beta}, \vec{\gamma})} &= U_{\text{QAOA}}(\vec{\beta}, \vec{\gamma}) \ket{+ \cdots +} \\
  &= \prod_{i = 1}^p e^{-i \beta_i \sum_{j=1}^n X_j} e^{-i \gamma_i H_{\text{c}}} \ket{+ \cdots +}, 
\end{split}
\end{align}
where the real vectors $\vec{\beta} = (\beta_1,...,\beta_p)$ and $\vec{\gamma} = (\gamma_1,...,\gamma_p)$ are variational parameters and $\ket{+ \cdots +}$ is the initial quantum state. By sampling from a quantum computer,  the expectation values $\bra{\psi(\vec{\beta}, \vec{\gamma})} H_{\text{c}} \ket{\psi(\vec{\beta}, \vec{\gamma})}$ can be computed and taken as the cost function for the optimization problem:
\begin{equation} \label{eq:qaoa_opt_prob}
    (\vec{\beta}^*, \vec{\gamma}^*) = \argmin_{\vec{\beta}, \vec{\gamma}} \bra{\psi(\vec{\beta}, \vec{\gamma})} H_{\text{c}} \ket{\psi(\vec{\beta}, \vec{\gamma})}.
\end{equation}

In practice, the optimization problem is solved in a variational manner with the usage of a classical optimization algorithm, yielding a quantum state $\ket{\psi(\vec{\beta}^*, \vec{\gamma}^*)}$ that approximates the ground state of $H_{\text{c}}$. This in turn allows the solution of the underlying QUBO problem to be extracted. Increasing the number of layers $p$ leads to an increase in the expressibility of the ansatz, and hence the potential quality of the output solution, at the expense of longer computation times \cite{zhou2020quantum}. We have chosen to investigate the original form of QAOA described in Ref. \cite{farhiqaoa} in our work, motivated by its relative simplicity and amenability to implementation on currently available quantum computers, with a few tens of qubits.

Next, we provide a general description of MCOPs. Given $L$ objective functions that map states from a state space to real numbers, multi-objective optimization asks for states that are optimal with respect to all of the $L$ objectives. Since the objectives generally produce competing effects with one another, states that are simultaneously optimal in all objectives generally do not exist. In this case, we ask instead for a set of Pareto-optimal/efficient states, which are optimal in the sense that no other states which improve on at least one individual objective without deteriorating in others can be found. On top of the objectives, a number of additional constraints may also be present, imposing further complications. 

More concretely, we require that the $L$ individual objectives and $K$ constraints of the problem can be cast into a QUBO problem. This yields an encoding of the states $x = (x_1,...,x_{n})$ as vectors of $n$ binary decision variables $x_1, ..., x_n \in \{0,1\}$ in a state space $S = \{0,1\}^n$, $L$ quadratic objective functions $C_i : S \rightarrow \mathbb{R}$ of the form:
\begin{equation}
    C_i(x) = x^\top \boldsymbol{Q_i} x, ~ i = 1,...,L,
\end{equation}
and $K$ quadratic penalty functions $P_j : S \rightarrow \mathbb{R}$ that enforce the $K$ constraints:
\begin{equation}
    P_j(x) = x^\top \boldsymbol{P_j} x, ~ j = 1,...,K,
\end{equation}
where the $\boldsymbol{Q_i}$'s and $\boldsymbol{P_j}$'s are upper triangular matrices with real entries. To compute these functions with quantum computers, we write the objective and penalty functions in terms of Ising variables $s = (s_1,...,s_{n}) \in S' = \{-1,1\}^n$, with $s_i = 2x_i - 1 \in \{-1, 1\}$, to eventually convert them into equivalent Hamiltonians, and denote the corresponding objective and penalty functions in terms of $s$ as $E^C_i(s)$ and $E^P_j(s)$ respectively. The MCOP is then:
\begin{equation} \label{eq:moop}
    \min_{s \in S'} ~ (E^C_1(s),..., E^C_L(s)),
\end{equation}
subject to the minimization of $E^P_1(s),..., E^P_K(s)$. 

Numerous classical methods exist to solve the above optimization problem. Of particular interest to us is linear scalarization, which seeks for a solution of Eq.~(\ref{eq:moop}) by first solving a simpler problem obtained by aggregating the objective and penalty functions of the full MCOP in a linear manner:
\begin{equation} \label{eq:moop_scalar}
    \min_{s \in S'} ~ \left( \sum_{i=1}^L w_i E^C_i(s) + w_P \sum_{j=1}^K E^P_j(s) \right),
\end{equation}
where the scalarization weights $w_i \geq 0$, and the penalty weight $w_P >0$ controls the degree at which infeasible solutions are penalized. A choice of $w_P$ that is sufficiently large ensures that the solution of Eq.~(\ref{eq:moop_scalar}) is necessarily a Pareto-optimal solution of Eq.~(\ref{eq:moop}) that satisfies the $P$ constraints imposed (i.e. a feasible solution) \cite{emmerich2018tutorial,krauss2020solving,hauke2021dominant}. In practice, smaller values can be used, and we set $w_P = 1$ throughout this work in an empirical manner, and set $w_i\in [0,1]$ without loss of generality (since only the relative weight $w_P/w_i$ matters).

For further details and references on MCOPs and scalarization, we refer the reader to Appendix.~\ref{appendix:moop}.

\subsection{Obtaining the Pareto Front with the QAOA} \label{subsec:obtaining}
\begin{figure*}[ht]
    \centering
    \includegraphics[width=\textwidth]{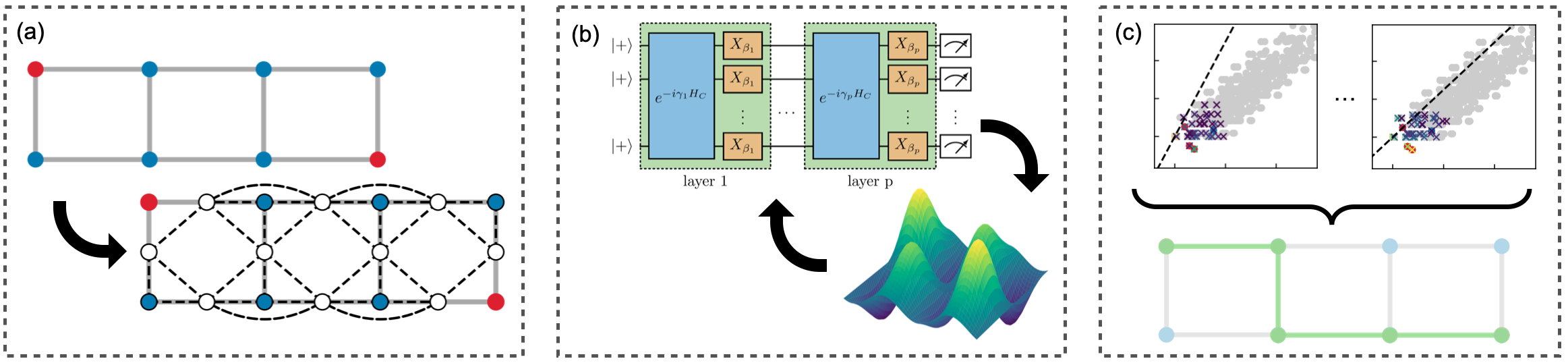}
    \caption{Summary of workflow to approximate the Pareto front of a MCOP. (a) Visualization of the conversion from the problem's underlying graph to the Hamiltonian corresponding to the QUBO formulation of the MCOP. The graph (with solid lines denoting its edges and circles denoting its nodes) is converted to the Hamiltonian Eq.~(\ref{eq:scalarized_qubo_ham}) that incorporates the objective and constraints of the problem. The Hamiltonian itself can be represented by a graph, $G_H$ (bottom right graph, with dotted lines with dotted lines denoting edges and circles denoting qubits). The illustration describes the conversion for the multi-objective routing problem, which is elaborated in Section.~\ref{sec:network_routing}. (b) Quantum-classical optimization with QAOA to solve the optimization problem Eq.~(\ref{eq:qaoa_opt_prob}) for the scalarized cost Hamiltonian Eq.~(\ref{eq:scalarized_qubo_ham}). The $p$-layer QAOA circuit is run at different angles according to the feedback of a classical optimizer, navigating the cost landscape until it converges to a minima. (c) Aggregation over results of one or more scalarization choices, determination of Pareto front with classical solver, and extraction of final result(s).}
    \label{fig:pareto_algo}
\end{figure*}

We now describe our procedure for solving MCOPs with near-term quantum computers, which is summarized in Fig.~(\ref{fig:pareto_algo}).

Following  Eq.~(\ref{eq:moop_scalar}), the QUBO objective vector $\vec{E}^C(s) = (E^C_1(s),..., E^C_L(s))$ is first scalarized by aggregating them as a convex sum with weights $\vec{w} = (w_1,..., w_L)$ with $w_i \in [0,1]$ and $\sum_{i=1}^L w_i = 1$. This is added to the penalty terms $\sum_{i=1}^K E^P_i(s)$ with each penalty function weighted equally by 1, resulting in a scalarized QUBO cost function:
\begin{equation} \label{eq:scalar_qubo}
    E_{\text{scalar}}(s) = \vec{w} \cdot \vec{E}^C(s) + \sum_{i=1}^K E^P_i(s).
\end{equation}
Minimization of this cost function then yields a feasible solution that corresponds to a Pareto-optimal solution $s^*$ of the MCOP Eq.~(\ref{eq:moop}). This optimization can be performed on a quantum computer by constructing equivalent objective Hamiltonians $H^C_1, ..., H^C_L$ and penalty Hamiltonians $H_1^P, ..., H_K^P$ via transformation to Ising variables, and aggregating them to result in an anologous scalarized cost Hamiltonian:
\begin{equation} \label{eq:scalarized_qubo_ham}
    H_{\text{scalar}} = \vec{w} \cdot \vec{H}^C + \sum_{i=1}^K H^P_i,
\end{equation}
whose ground state encodes the same solution $s^*$, where $\vec{H}^C = (H^C_1, ..., H^C_L)$. We perform the ground state search with a VQA, chosen to be QAOA in our work due to its simplicity and amenability by current quantum computers. That is, we variationally optimize the QAOA circuit of Eq.~(\ref{eq:qaoa_unitary}) to produce an optimized output quantum state $\ket{\psi(\vec{\beta}^*, \vec{\gamma}^*)}$ that approximates the ground state of $H_{\text{scalar}}$, which is then sampled from $k$ times to yield a set of candidate solutions $\bm{B} = \left\{b_1,...,b_k\right\}$, with $k$ always chosen to be at most polynomial in the number of qubits $n$. The set $\bm{B}$ then constitutes a reduced search space within which Pareto-optimality can be efficiently checked with any classical method (such as brute force). The output of this procedure is therefore expected to be a set of feasible solutions that are Pareto-optimal within $\bm{B}$. As the number of QAOA layers is increased to $\infty$, the QAOA ansatz supports an increasingly better approximation of the ground state of Eq. \ref{eq:scalarized_qubo_ham}. Provided that it can be found in the variational procedure, sampling this state would then return solutions that are also Pareto-optimal within $S$, with high probability.

The above scalarization procedure involves the linear aggregation of objectives. The convex weights $w_i$ can be interpreted as \textit{a priori} preferences that the decision maker can select, which biases the resulting solution according to $\vec{w}$. While other scalarization schemes that involve higher order terms (such as quadratic scalarization) and inequality constraints (such as Chebyshev scalarization) are available, we focus on the linear scheme, as it preserves the QUBO form of the individual objectives and constraints. The resulting scalarized cost function Eq.~(\ref{eq:scalar_qubo}) then contains only linear and quadratic terms. We discuss possible limitations and extensions of this approach in Appendix.~\ref{appendix:moop}.

So far, this procedure belongs to the class of \textit{a priori} methods, where the decision makers' preferences are specified before the optimization to bias the output of the optimization procedure (c.f. Appendix.~\ref{appendix:moop} for details). To recover the entire Pareto front in an unbiased way, this procedure can be converted into a completely \textit{a posteriori} method by repeating the procedure with numerous choices of $\vec{w}$ in a problem-agnostic manner, with e.g. uniform random sampling of the weights or a discretization over all possible weights. The full procedure to recover the Pareto front with QAOA is illustrated in Fig.~(\ref{fig:pareto_algo}) -- $M$ different scalarization weights are chosen corresponding to $M$ QAOA runs, resulting in $M$ sets of candidate solutions $\bm{B}^1,..,\bm{B}^M$ which are aggregated as $\bigcup\limits_{i=1}\limits^{M}\bm{B}^i$. A classical method then checks for Pareto-optimality within this set, yielding a set of feasible solutions that approximates the Pareto front.

\subsection{Visualization of Solutions in the Pareto Plot}
\label{subsec:Pareto_plots}
The state space of an $L-$objective MCOP can be visualized in a Pareto plot \cite{boyd2004convex,emmerich2018tutorial}, which is an $L-$dimensional plot with $2^{|S|}$ points. Each point corresponds to a state/bitstring $s$, with the value of the $L$ objectives as its coordinates. Pareto-optimal solutions then constitute points located at the boundary of the region populated by the $2^{|S|}$ points \cite{boyd2004convex}. 

In our case, we must also account for the penalty terms $E^P_1(s),...,E^P_K(s)$, whose contribution should be reflected in the Pareto plot in a consistent way. This can be achieved by defining a point $\vec{r} = (r_1,...,r_L)$ in the $L$-dimensional Pareto plot as follows:
\begin{equation} \label{eq:projection defn}
    r_i(s) = \sum_j^K E^P_j(s) + E^C_i(s), ~~ i = 1,...,L.
\end{equation}
This definition has the property that the cost function arising from linear scalarization with weights $\vec{w}$ can be interpreted as the projection of points in the Pareto plot onto the vector $\vec{w}$, since:
\begin{equation}
    \vec{r} \cdot \vec{w} = \sum_j E^P_j(s) + \vec{w} \cdot \vec{E^C}(s),
\end{equation}
which is exactly the scalarized cost function Eq.~(\ref{eq:scalar_qubo}) that is minimized in our procedure.

We now give a brief qualitative discussion of some notable features of the Pareto plot obtained from the application of QAOA to a MCOP, chosen to be a small network routing problem (to be defined in Section.~\ref{sec:network_routing}) defined on 13 qubits for illustration. While the precise details of a Pareto plot will naturally differ depending on the specific problem instance, the observations we make in the following were general to all the cases we have considered in this work.

\begin{figure*}
    \centerline{\includegraphics[width=1\textwidth]{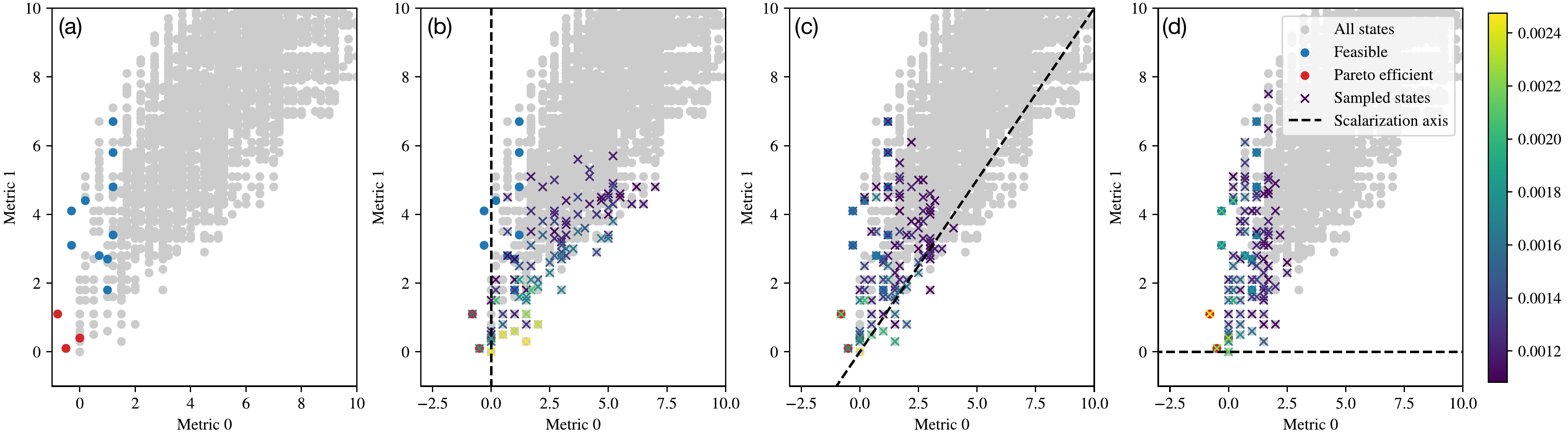}}
    \caption{Pareto plot of a 4-objective problem, projected on two of the metrics. Each point in the plot corresponds to a possible state/solution (for a total of $2^{n_{\text{qubit}}}$ points), with feasible solutions marked as blue and Pareto-optimal solutions marked as red. We also select the top $k = 100$ most probable states from $\ket{\psi(\vec{\beta}^*, \vec{\gamma}^*)}$ and color code them according to their probability (colored crosses). We plot the results for 3 different scalarization choices, with the gradient of the scalarization weight vector visualized (black dotted lines passing through origin). The scalarization vectors are $\vec{w} = (0,1,0,0)$ for (a), $\vec{w} = (1/4,1/4,1/4,1/4)$ for (b), and $\vec{w} = (1,0,0,0)$ for (c). The underlying network is a triangular lattice graph with 6 nodes and 9 edges, requiring 13 qubits to encode.}
    \label{fig:pareto_qaoa}
\end{figure*}

Fig. \ref{fig:pareto_qaoa}.(a) shows the Pareto plot for our 13-qubit network routing problem (with parameters specified in Table.~\ref{appendix:fig_problem}). The problem involves 4 objective functions, and the figures show a projection onto the plane of two of those objectives. Each point in the plot corresponds to a possible state (a bitstring), with feasible states marked as blue and Pareto-optimal solutions marked as red. The top-$k$ most probable states from $\ket{\psi(\vec{\beta}^*, \vec{\gamma}^*)}$ are marked with crosses and color coded according to their probability (colored crosses). In Fig. \ref{fig:pareto_qaoa}.(b),(c),(d) we illustrate the solution returned by QAOA with different scalarization weights ($\vec{w} = (0,1,0,0)$ for (a), $\vec{w} = (1/4,1/4,1/4,1/4)$ for (b), and $\vec{w} = (1,0,0,0)$ for (c)).

A few important observations from these plots are the following:
\begin{itemize}

    

    \item \textbf{The states with highest sampled probability in the output state $\ket{\psi(\vec{\beta}^*, \vec{\gamma}^*)}$ are in the low-energy sector of the Pareto plot} : 
    
    This is a direct consequence of the quantum-classical optimization loop that results in an output quantum state $\ket{\psi(\vec{\beta}^*, \vec{\gamma}^*)}$ that has minimal scalarized energy when optimized. Sampling from $\ket{\psi(\vec{\beta}^*, \vec{\gamma}^*)}$ then leads to a set of candidate states $\bm{B}^i$ that are biased to contain high-quality/low-energy states of the scalarized problem, which we can observe from the clustering of the colored crosses in the bottom left region of the Pareto plot.
    
    \item \textbf{States sampled from $\ket{\psi(\vec{\beta}^*, \vec{\gamma}^*)}$ are squeezed in the direction of the scalarization line} : 
    
    Scalarization amounts to projecting points on the Pareto plot to the line passing through the origin with gradient equal to the scalarization weight vector (black dotted line). Different choices of scalarization therefore result in different regions in the Pareto plot that are favoured. This feature can be observed clearly through the Pareto plots by the clustering of high probability states (yellow, light colored crosses) along the direction of the scalarization line. As an example, for the scalarization choice corresponding to weighting Metric 2 with unit weight (second of the four plots in Fig.~(\ref{fig:pareto_qaoa})), the sampled states can be observed to be squeezed vertically, towards the bottom of the Pareto plot. 
    
    \item \textbf{Points in the Pareto front corresponding to solutions of other scalarization choices can be sampled} : 

    If there are other Pareto-optimal states close to the solution of the scalarized problem in the Pareto plot, in the sense that they are low-energy states of the scalarized problem, the set of candidate states $\bm{B}^i$ sampled from $\ket{\psi(\vec{\beta}^*, \vec{\gamma}^*)}$ may contain these nearby Pareto-optimal solutions with relatively high probability. In principle, this allows multiple Pareto-optimal states to be obtained from one iteration of the procedure. For example, in Fig.~(\ref{fig:pareto_qaoa}) we observe that the Pareto-optimal solution corresponding to the rightmost red point is sampled with relatively high probability in all three scalarization choices, even though it is not a solution of any of these scalarized Hamiltonians. We observe that this is often the case in problem instances we consider.
    
    \item \textbf{Points in the Pareto front located in locally concave regions can be sampled} : 
    
    Again, because states near the target Pareto optimal point can be sampled with high probability from $\ket{\psi(\vec{\beta}^*, \vec{\gamma}^*)}$, states located at locally concave regions of the Pareto front can in principle be sampled as long as they are near the target (in the Pareto/objective space), even with linear scalarization. This allows Pareto-optimal points located at concave regions to be identified, even though they can never correspond to solutions of scalarized cost functions. Using the rightmost red point in Fig.~(\ref{fig:pareto_qaoa}) as an example again, we observe that it is sampled with relatively high probability, despite being located at a locally concave region, which cannot be obtained as the solution of any scalarized Hamiltonian.
    
\end{itemize}

The latter two observations indicate a degree of robustness to the choice of scalarization and the concavity of the Pareto region, due to QAOA being a sampling-based algorithm, which requires sampling from the optimized quantum state in order to extract information it. This introduces a trade-off between the ability of QAOA to approximate the ground state (by increasing $p$), and for $\ket{\psi(\vec{\beta}^*, \vec{\gamma}^*)}$ to remain as a superposition of energy eigenstates so that low-energy states can be sampled (by keeping $p$ small), potentially imposing a less stringent scaling of $p$ with problem size.

Finally, we remark that most points in the Pareto plots correspond to infeasible solutions, which violate at least one of the problem constraints encoded in the penalty terms of Eq.~(\ref{eq:scalarized_qubo_ham}). As we explain in Appendix.~\ref{appendix:Feasible_Infeasible}, in general the fraction of feasible to infeasible solutions decreases rapidly with the network size. This fact represents a challenge for QUBO approaches where the encoding of constraints as penalty terms is unable to restrict the search through solutions to remain solely within the feasible subspace. We discuss this point further in Section.~\ref{subsec:QAOA_mixer_initial_state} in the specific context of QAOA.

\section{Application to the Network Routing Problem} \label{sec:network_routing}
With a general description of our algorithm for MCOPs established, we now consider the concrete example of the multi-objective network routing problem. 

We consider a generic multi-hop wireless network with relay stations distributed throughout a geographical area. This can be modelled as an undirected weighted graph $G = (V(G), E(G))$ (which we also refer to as the network's graph) with nodes $V(G)$ that correspond to relay stations and edges $E(G)$ defined by possible transmission paths between stations (Directed graphs can be included with a more general formulation \cite{krauss2020solving}). Two nodes are specified to be the source and destination nodes (denoted $s$ and $d$ respectively). The routing problem then asks for data transmission routes between $s$ and $d$ that are Pareto optimal with respect to multiple objectives, which can be written as functions of the node and edge weights of the graph. This is an instance of the multi-objective shortest path problem, which is a combinatorial optimization problem known to be NP-hard \cite{serafini1987some}. Its solution is relevant across a wide range of network design and optimization tasks \cite{akyildiz2002survey,pantazis2007survey,kulkarni2010computational,sarangapani2017wireless,bockelmann2018towards,sharma2019toward}, especially beyond current wireless network protocols based on individual objectives such as the Optimized Link State Routing protocol \cite{jacquet2001optimized} which only considers hop count, and is thus unable to maximize network resource utilization. 

The form of the objectives depends on specific performance demand requirements. We consider the following four objectives in our work:
\begin{enumerate}
\item \textbf{Path loss}: Transmission between two stations incurs an energy cost that is dependent on their physical distance.
\item \textbf{Node Delay}: Signal processing at each station incurs a time delay, which sums up to an overall delay.
\item \textbf{Data rate}: The total data output rate of a transmission path is determined by the minimum data rate along its path.
\item \textbf{Bit error}: Bit errors occur with finite probability during transmission between two stations, which depends on the path loss and the channel's noise profile.
\end{enumerate}

To solve this problem with our approach, a QUBO encoding of the objectives and constraints of the multi-objective shortest-path problem is needed. We provide a complete description of this encoding in terms of binary variables together with the form of the quadratic cost function in Appendix.~\ref{appendix:shortest} (following \cite{krauss2020solving}). This is followed by QUBO formulations of the above four objectives in Appendix.~\ref{appendix:network}. Along with reasonable assumptions on network parameters based on software-defined radio use cases \cite{alanis2014quantum}, the associated objective and penalty Hamiltonians $H^C_i$ and $H^P_i$ can then be constructed explicitly, thereby fully specifying the inputs for our procedure in Fig.~(\ref{fig:pareto_algo}).

Importantly, we observe that for a network with graph $G$, the graph of its corresponding scalarized Hamiltonian $G_H$ -- that is, the graph with connectivity defined by the quadratic terms of Eq.~(\ref{eq:scalarized_qubo_ham}) -- can be interpreted as the network graph's \textit{middle graph} \cite{hamada1976traversability}, denoted as $M(G)$. Using known properties of $M(G)$, we will exploit this correspondence in the next sections to bound the resources needed to execute the algorithm.  The Hamiltonian's construction is illustrated in Fig.~(\ref{fig:pareto_algo}).(a). The network's graph $G$ (top left graph, with solid lines denoting its connectivity, red circles denoting source and destination nodes, and blue circles denoting remaining nodes) is converted to a Hamiltonian $H$, with connectivity defined by the graph $G_H$, which takes the form of $M(G)$ (bottom right graph, with dotted lines denoting the connectivity of $G_H$ or $M(G)$, and circles denoting nodes). We elaborate on points regarding $M(G)$, including its definition and properties, in Appendix.~\ref{appendix:graph_theory}.

\section{Theoretical and Numerical Scaling Analyses} \label{sec:analyses}

We now turn to analyze in detail the scalability and performance of the QAOA approach to the network routing problem.

In Section \ref{subsec:QAOA_mixer_initial_state}, we begin with a short qualitative discussion of the role of the QAOA mixer and initial state in determining the efficacy of the algorithm. Subsequently, in Section \ref{QAOA_Resource_Estimates}, we discuss and provide a pragmatic analysis of the hardware resources required to implement a problem instance of given size, and compare these requirements against known limitations of VQAs. This allows us to distinguish between problem instances that are infeasible for our approach, from those that are potentially amenable. Finally, in Section \ref{subsec:numerical} we present the results of numerical simulations that explore the scaling of several success metrics as a function of problem size and the QAOA depth $p$. We remark that due to limitations on the system sizes that can be simulated on a classical computer, strong conclusions on the asymptotic performance of the algorithm cannot be drawn at this stage.

All analyses performed here are based on a single scalarization (i.e. we apply the algorithm in an \textit{a priori} manner by pre-specifying the scalarization weights $\vec{w}$). As described in Section.~\ref{subsec:obtaining}, if the intent is to recover the entire Pareto front instead, the total runtime of the procedure depends on the number of scalarizations, which introduces an additional multiplicative overhead in the time required to obtain the Pareto front. As is also the case with classical approaches, this additional complexity is highly problem-dependent.

\subsection{QAOA mixer and initial state}\label{subsec:QAOA_mixer_initial_state}

We begin with a qualitative discussion of the role of the QAOA mixer and circuit initial state in determining the efficacy of the approach. In this work, we have used the standard QAOA initial state and mixer pair \cite{farhiqaoa}, where the circuit is initially prepared in an equal superposition of all solutions, i.e. $|+\rangle^{\otimes n}$, and the mixer Hamiltonian $H_{m} = -\sum_i^n X_i$ drives bit flips across the register. Constraints are enforced through energy penalties in the cost Hamiltonian Eq.~(\ref{eq:scalarized_qubo_ham}), with the search ideally converging towards low-energy feasible solutions. However, the solution space is increasingly dominated by infeasible configurations as the network size grows (see Appendix.~\ref{appendix:Feasible_Infeasible}), motivating the need for alternative ways of searching the solution space.

In the context of the network routing problem, we leave the question of designing improved initial state and mixer Hamiltonian pairings for future work (see also our remarks in Section.~\ref{subsec:qaoa_outlook}). However, we note that the goal of such strategies is to reduce the size of the solution space to be searched, either by entirely avoiding infeasible solutions, or by avoiding some subset of them. In the former case, where the search takes place through feasible solutions only, the penalty terms in Eq.~(\ref{eq:scalarized_qubo_ham}) can be eliminated. In the latter case (which may arise if a strategy to search only feasible solutions cannot be found, or carries impractical resource requirements) the penalty terms would still be necessary to enforce constraints indirectly (i.e. through the objective function)

With these considerations in mind, in the following subsections we give a detailed analysis of the resource requirements of Eq.~(\ref{eq:scalarized_qubo_ham}). We also remark that in the standard QAOA approach (Eq.~(\ref{eq:qaoa_unitary})), the initial state $|+\rangle^{\otimes n}$ can be easily prepared by applying a Hadamard gate to each qubit. Since the mixer Hamiltonian $e^{-i \beta_i \sum_{j=1}^n X_j}$ also involves only single-qubit operations, its implementation is also easy.

\subsection{Cost Hamiltonian: Theoretical Resource Estimations} \label{QAOA_Resource_Estimates}

To determine the resource requirements of our QAOA approach, we examine the runtime and number of qubits necessary for its implementation on quantum hardware. This is determined by considering the resources needed to implement each layer of the QAOA circuit, and the number of repeated executions required to estimate the cost function up to an error $\epsilon$. 




We remark that for a given error tolerance $\epsilon$, a complete performance analysis would involve exposing the dependence of the number of required QAOA layers $p$ on problem size, which is challenging from both an analytic and numerical perspective. This is further related to the issue of trainability and existence of barren plateaus of VQAs \cite{NIBP_paper,mcclean2018barren,cerezo2021cost}, which depend strongly on problem class, presence of noise, and ansatz choice, and is a subject of ongoing study. Specifically, if the physical circuit depth for an application scales super-linearly in $n$, the optimization procedure will suffer from a \emph{noise-induced barren plateau} \cite{NIBP_paper}, implying expensive gradient computations that scale exponentially in $n$, a phenomenon which is conceptually similar to the issue of vanishing gradients that previously plagued the training of neural networks in classical machine learning \cite{ML_vanishing_gradients}.


We begin by outlining a few assumptions and simplifications that we will make. Firstly, the quantum circuit of a $p$-layer QAOA consists of $p$ alternating mixer and cost unitaries (Eq.~(\ref{eq:qaoa_unitary})). Since elementary gates in the cost unitary $e^{-i \gamma_i H_{\text{c}}}$ fully commute with one another, the 1-qubit RZ gates can be scheduled to be executed first, in parallel. Together with the fact that the mixer unitary consists only of 1-qubit RX gates which can also be executed simultaneously, and that 1-qubit gates execute significantly faster than 2-qubit gates, we only need to consider contributions from 2-qubit gates. Secondly, we neglect compilation overheads arising from qubit routing, i.e. the need to include additional SWAP gates to carry out 2-qubit gates that are not natively executable on a quantum processor. While this is a valid assumption for quantum processors with a fully-connected topology (such as currently available ion trap devices, which we utilise in Section.~\ref{sec:qpu_results}), the mismatch between the quantum processor and the cost Hamiltonian's topologies will generally require routing. The routed output depends on the degree of mismatch and the routing strategy employed, among many other factors which are beyond the scope of our discussion. Nonetheless, we remark that the inclusion of $n-1$ additional layers of SWAP gates is a naive upper bound \cite{kivlichan2018quantum,o2019generalized}, incurring at most $O(n)$ additional layers of quantum gates. Thirdly, to streamline our argument we ignore quadratic terms arising from the source and destinations constraints, which only incurs a negligible, constant number of terms (detailed in Appendix.~\ref{appendix:shortest}). 

Finally, we remark that in our formulation of the problem, increasing the number of objectives solely results in additional linear terms in the cost Hamiltonian Eq.~(\ref{eq:scalarized_qubo_ham}), since only the penalties contain quadratic terms. An increase in the number of objectives therefore does not directly incur additional time and qubit costs. 

\subsubsection{Number of qubits}
Since the graph of the cost Hamiltonian $G_H$ can be mapped
to the middle graph $M(G)$, the number of qubits involved is equal to the number of nodes of $M(G)$, resulting in $n = |V(G)| + |E(G)| = O(|V(G)| \Delta_G)$, where $\Delta_G$ is the maximum degree of the network's graph $G$ (see Appendix.~\ref{appendix:graph_theory}, Eq.~(\ref{eq:mg_nodes})). This implies that problem instances with maximum degree independent of problem size are expected to be more efficient, with $n = O(|V(G)|)$.

\subsubsection{Cost computation time}
The computation time of the $p$-layer QAOA cost function is determined by two factors: (1) the number of repetitions $n_{\text{rep}}$ required to estimate the cost function expectation value, given some specified error tolerance, and; (2) the depth of each executed circuit, which we denote $D_p$ for $p$ layers. These two factors combine to give an execution time of $O(D_p n_{\text{rep}})$, or $O(p D_1  n_{\text{rep}})$ in terms of the depth of a single QAOA layer $D_1$.

$D_1$ depends on the number and connectivity of the 2-qubit terms in the cost Hamiltonian. As they are Ising gates that are diagonal in the computational basis, they fully commute with one another, and can be scheduled to maximize parallelization. The determination of such a schedule amounts to the solution of an edge-coloring problem on $G_H$, where edges (i.e. gates) of the same color are scheduled to be executed simultaneously. The minimal number of colors required is commonly termed the edge chromatic number $\chi'(G_H)$, so we have $D_1 = \chi'(G_H)$ layers in total. To estimate $\chi'(G_H)$, note that the structure of $G_H$ is completely determined by Eqs.~(\ref{source_constraint})-(\ref{path_constraint}). Intuitively, we expect $G_H$ to inherit the local structure of $G$, since the squares in Eqs.~(\ref{source_constraint})-(\ref{path_constraint}) only lead to interactions between nodes (of $G_H$) representing nodes and edges incident to them (i.e. the $x_i x_{ij}$ terms), and between edges that are incident on the same node (i.e. the $x_{ij} x_{ik}$ terms). It turns out that this is precisely the relationship between a graph $G$ and its middle graph $M(G)$ -- in other words, $G_H = M(G)$, and hence $\chi'(G_H) = \chi'(M(G))$. This realization allows us to leverage on known properties of $M(G)$ to show that:
\begin{equation} \label{eq:edgecolor_bound}
    \chi'(M(G)) \leq 2 \Delta_G,
\end{equation}
where we delegate the proof and a review of relevant ideas to Appendix.~\ref{appendix:graph_theory}. We conclude that the depth $D_1$ of each QAOA layer scales linearly with the maximum degree of the network's graph $G$, i.e. $D_1 = O(\Delta_G)$.

On the other hand, it is known that $n_{\text{rep}} = O(L/\epsilon^2)$ measurements are needed to compute expectation values up to an error tolerance of $\epsilon$ by operator averaging \cite{peruzzo2014variational}, where $L$ is equal to the number of Pauli terms in the operator (which is the cost Hamiltonian in our case). Of the $L$ terms, $|V(G)| + |E(G)| \leq \frac{1}{2}|V(G)|(2 + \Delta_G)$ terms are linear (since the number of node and edge weights is equal to the total number of nodes and edges of $G$), while $|E^{M(G)}| \leq \frac{1}{2}|V(G)|\Delta^2_G$ terms are quadratic (from Eq.~(\ref{eq:mg_edges}), because $G_H = M(G)$), for a total of $L \leq  \frac{1}{2}|V(G)|(2 + \Delta_G + \Delta_G^2)$ terms. Thus, $n_{\text{rep}}$ scales asymptotically as $O(|V(G)| \Delta_G^2/ \epsilon^2)$.

Summarizing, the computation time of a QAOA cost function up to an error $\epsilon$ scales as $O(p |V(G)| \Delta_G^3 / \epsilon^2)$, which can be parallelized up to a factor of $n_{\text{rep}}$.

We remark that this implies efficiency for problem classes with a connectivity or maximum degree that is independent of the problem size (such as $k$-regular/lattice-type graphs), with only a linear dependence of the computation time in problem size. This is likely the scenario for large-scale applications, since the cost of supporting a dense network (such as fully-connected mesh networks) over large geographical regions will otherwise be high, rendering it practically infeasible. In that case, the computation time (and the number of qubits needed) becomes at most quadratic in problem size.


\subsection{Numerical Scaling Analyses} \label{subsec:numerical}

\subsubsection{Performance Metrics}
We now introduce several measures that quantify the degree to which our scheme has successfully solved the task at hand, which is to obtain Pareto-optimal points of a multi-objective problem.

A standard measure of the quality of solution output by QAOA is the \textit{approximation ratio}, defined as the ratio between the energy of the output state and the ground state:
\begin{equation}
    r_{\mathrm{approx}} \coloneqq \frac{\bra{\psi(\vec{\beta}^*, \vec{\gamma}^*)} H \ket{\psi(\vec{\beta}^*, \vec{\gamma}^*)}}{E_{\text{min}}}.
\end{equation}
An approximation ratio of 1 thus implies that QAOA has found the exact ground state of $H$.

In our multi-objective context, however, we ask for the set of Pareto-optimal solutions, which generally cannot be encoded as ground states of a single Hamiltonian. As a more pragmatic measure in a multi-objective context, we supplement analyses of $r_{\mathrm{approx}}$ by defining the \textit{success probability}, which is the probability that a Pareto-optimal state is sampled from $\ket{\psi(\vec{\beta}^*, \vec{\gamma}^*)}$,
\begin{equation}
    p_{\mathrm{success}} \coloneqq \sum_{x_i \in \mathrm{Pareto~front}} |\langle \psi(\vec{\beta}^*, \vec{\gamma}^*) | x_i \rangle|^2.
\end{equation}
A success probability of 1 implies that sampling from $\ket{\psi(\vec{\beta}^*, \vec{\gamma}^*)}$ will always yield a Pareto-optimal state. Note that it does not contain information about whether it contains points from the entire Pareto front. 

\subsubsection{Scaling with System Size and Number of Layers}\label{QAOA_layers_scaling}
In this section, we present results from classical simulations of QAOA to study the scalability of the approach with respect to different problem sizes, graph geometries, and the number of QAOA layers $p$. These simulations assume an absence of both statistical and hardware noise. Taking into account limitations on the classical simulation and optimization of general QAOA circuits for large $n$ and $p$, and the need to average over multiple problem instances, we consider problems of sizes up to 16 qubits and $p=10$ layers. 

\begin{figure*}[ht]
    \centering
    \centerline{\includegraphics[width=1.5\columnwidth]{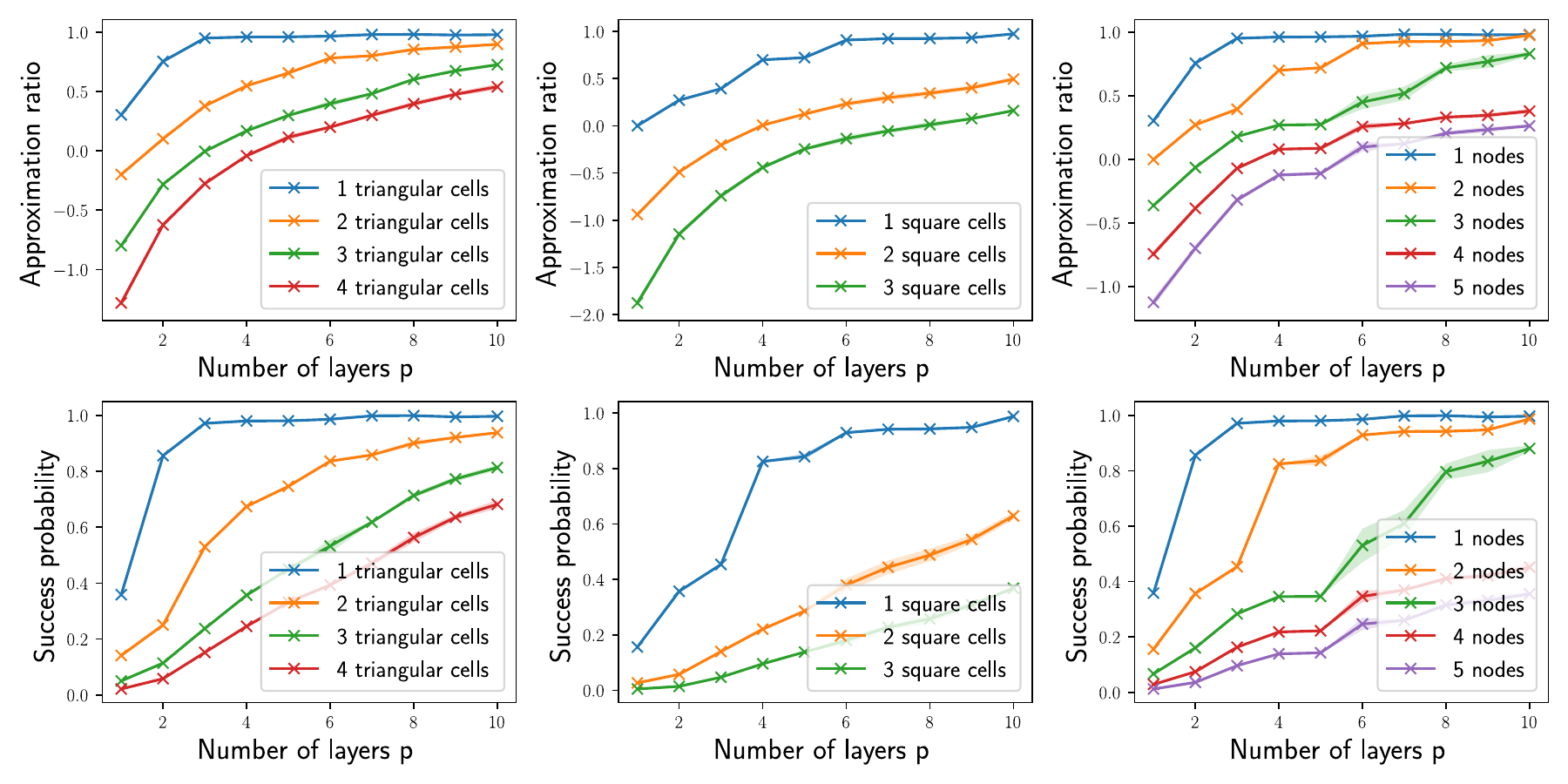}}
    \caption{Scaling of the approximation ratio and success probability with the number of layers for triangular lattices, square lattices, and cycle graphs (left to right columns). Different colors indicate different problem sizes, with standard deviations displayed in shaded regions. For cycle graphs (rightmost column), the legend indicates the number of nodes besides the soure and destination nodes. The data was produced after optimization with linear ramp initialization, two metrics (node based and edge based respectively), with node/edge weights uniformly randomized between -1 and 1, averaged over 50 instances. scalarization weights are chosen to be equal for both objectives, i.e $\vec{w} = (0.5, 0.5)$.}
    \label{fig:scaling}
\end{figure*}

Fig.~(\ref{fig:scaling}) displays the scaling of the approximation ratio and success probability as a function of $p$, for different problem sizes chosen from different graph geometries, namely triangular lattices, square lattices, and cycle graphs. For all examples, we consider two linear objectives, one involving only edge variables (i.e. of the type Eq.~(\ref{eq:edge_cost})) and another only involving node variables (i.e. of the type Eq.~(\ref{eq:node_cost})), with node/edge weights uniformly randomized between -1 and 1, averaged over 50 instances. The scalarization weights are also chosen to be equal for both objectives, i.e $\vec{w} = (0.5, 0.5)$. Finally, initial parameters for QAOA are chosen according to a linear ramp initialization scheme, which is a heuristic choice that linearly ramps up $\gamma$'s and ramps down $\beta$'s, based on the analogy between QAOA and quantum annealing \cite{zhou2020quantum}. Additional numerical results for the performance of this initialization scheme can be found in Appendix.~(\ref{appendix:initializations}).

As first observations, we see that both metrics increase monotonically as a function of $p$. This is in accordance with the fact that as $p\rightarrow \infty$ with QAOA, we recover the limit of infinitesimally slow adiabatic quantum annealing, and recover the ground state/optimal solution of scalarized cost Hamiltonians with unit probability. For any fixed value of $p$, both metrics also achieve higher values for smaller problems, again mirroring the increase in annealing time needed for larger problems. Finally, we observe that problem classes with higher average connectivities tend to reach saturation at a slower rate (number of layers to reach saturation, in descending order : cycle graphs, triangular lattice, square lattice). These conclusions hold for generic problem instances of the same class, due to the average over numerous instances.

Focusing on the success probability (lower row of Fig.~(\ref{fig:scaling})), we find that it increases monotonically with $p$, with indications of saturation at the maximum value of 1 for the limited problem sizes we have considered. These results suggest that, at least for the problem instances considered here, the procedure indeed outputs Pareto-optimal solutions with increasing frequency as we increase $p$. This justifies the pragmatic approach of increasing $p$ to improve the quality of the procedure.

Nonetheless, we note that the aforementioned difficulty in classically simulating and optimizing large QAOA circuits prevents further scaling conclusions to be drawn. We leave a more complete investigation of these issues, involving the large-scale benchmarking of our scheme on actual quantum computers, as further work. As a final remark, a positive indication is provided in the closely related context of quantum annealing, by the polynomially vanishing energy gap between the ground and first excited state of the Hamiltonian Eq.~(\ref{eq:scalarized_qubo_ham}) \cite{hauke2021dominant}, which is strongly correlated to a better performance of QAOA \cite{zhou2020quantum}.

\section{Computations on Quantum Computers} \label{sec:qpu_results}
In this section, we describe an implementation of our scheme for a small-scale network routing problem on the 11-qubit IonQ Harmony quantum computer, accessed through the Amazon Braket cloud quantum computing service with the OpenQAOA Python SDK \cite{sharma2022openqaoa}. These experiments serve to empirically verify that the scheme can yield Pareto-optimal solutions to the network routing problem, in a manner consistent with our description in Section.~(\ref{sec:qc_moop}). We consider a 4-node fully-connected network and a 6-node square-lattice network, requiring 8 and 11 qubits respectively to encode in QUBO form. For both problem instances, we consider 4 objectives relevant to the network routing problem: data rate, path loss, node delay, and bit-error-rate. Further details on the problem instances can be found in Appendix.~(\ref{appendix:problem_instances}).

We run the algorithm for the scalarization choice $\vec{w} = (1/4, 1/4, 1/4, 1/4)$, with $p=1$ and 2000 shots per circuit execution, and the standard QAOA circuit involving standard single-qubit $RX$ mixers and uniform computational basis state initialisation. Starting at initially suboptimal parameters, COBYLA \cite{Powell1994} is chosen as the classical optimization algorithm. 

\begin{figure}
\centering
\subfloat{\includegraphics[width=0.485\linewidth]{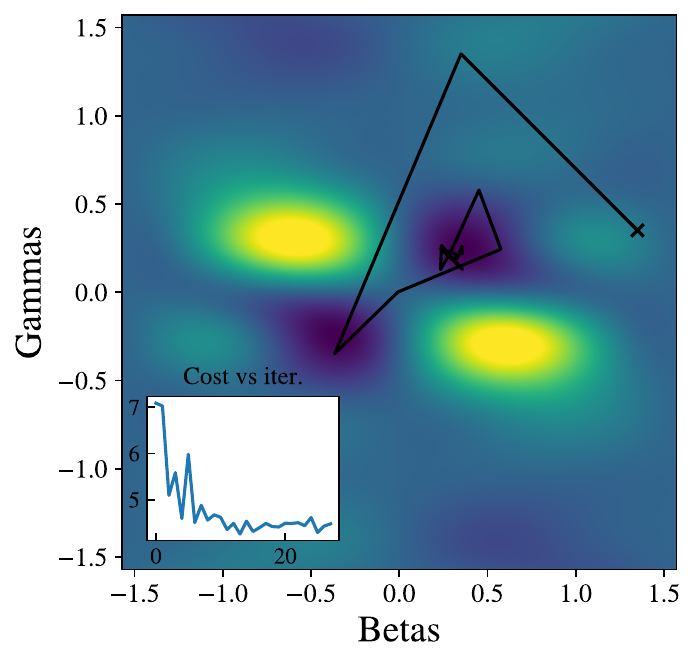}
    \label{fig:cost_first}}
\subfloat{\includegraphics[width=0.55\linewidth]{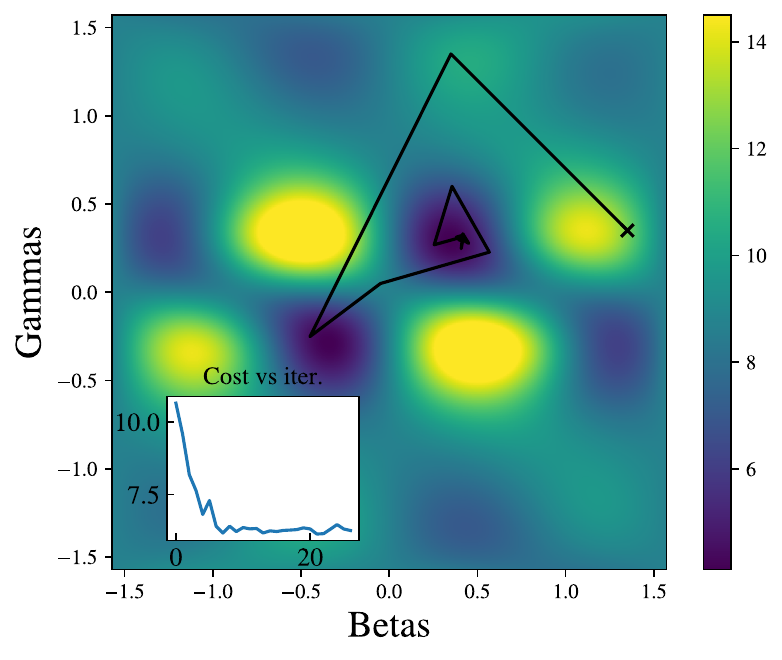}
    \label{fig:cost_second}}

\caption{Trajectory of the QAOA parameters during optimization (black line) for both the 8-qubit fully-connected network (left plot) and 11-qubit square lattice networks (right plot), overlaid on the full cost function landscape obtained by classical simulation of the QAOA circuits. Initial parameters are displayed as black crosses. Inset shows cost vs iteration plots of the optimization.}
\label{fig:qpu_cost_prog}
\end{figure}

The results of the optimization are shown in Fig.~(\ref{fig:qpu_cost_prog}), which displays the trajectory of the parameters $\beta$ and $\gamma$ (black line), plotted on top of the cost function landscape obtained by classical simulation. The inset at each plot shows the evolution of the cost function value during optimization. Starting from suboptimal initial parameters, we observe convergence to local minima for both problems. 

\begin{figure*}[t]
\centering
\subfloat[Fully-connected graph, 8 qubits]{\label{fig:8q_ionq}
\centering
\includegraphics[width=0.5\linewidth]{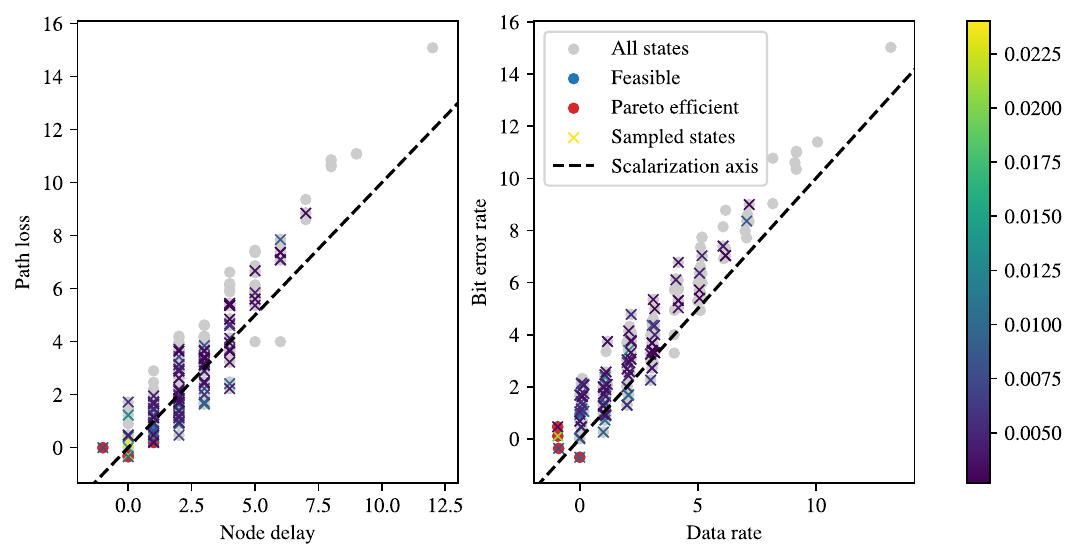}
}
\subfloat[Square lattice, 11 qubits]{\label{fig:11q_ionq}
\centering

\includegraphics[width=0.5\linewidth]{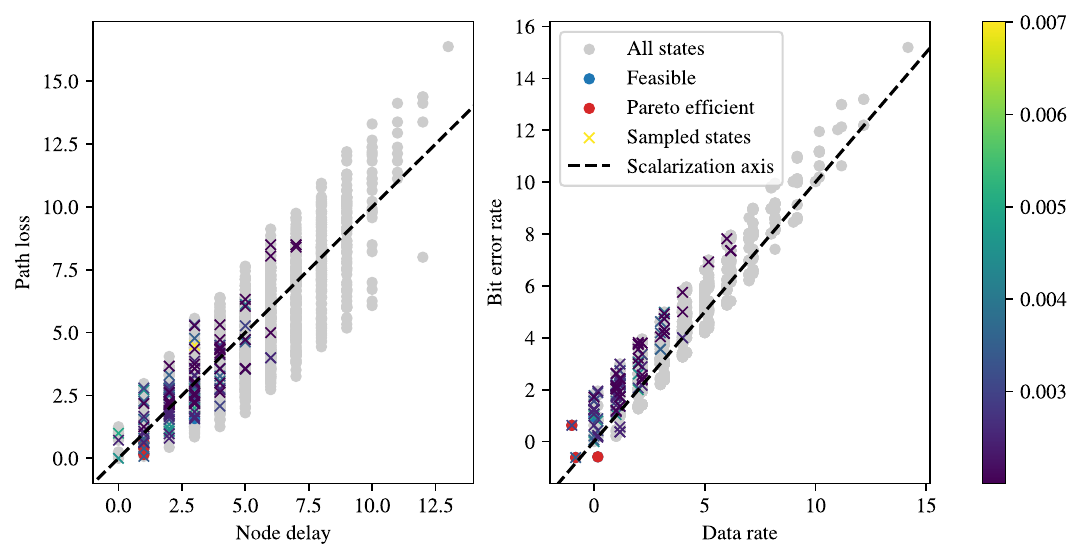}
}
\caption{Pareto plots obtained from sampling the optimized quantum state of QAOA with the IonQ device for both the 8-qubit fully-connected network (left two plots) and 11-qubit square lattice networks (right two plots). Each point in the plot corresponds to a possible state, with feasible solutions marked as blue and Pareto-optimal solutions marked as red. The top $k = 125$ most probable states sampled are plotted as crosses and color coded according to its probability. The gradient of the scalarisation weight vector is also visualised (black dotted lines).}
\label{fig:qpu_pp}
\end{figure*}

To obtain Pareto-optimal solutions of the network routing problem, we sample from the optimized quantum states. Fig.~(\ref{fig:qpu_pp}) displays the Pareto plot for two pairs of objectives obtained from the IonQ device, with the top $k=125$ most probable states marked as coloured crosses (we limit $k$ for visual clarity). Consistent with our discussions in Section.~(\ref{sec:qc_moop}), we observe the clustering of high-probability states close to the Pareto front in these plots. The observed success probabilities are also computed to be $\approx$~0.008 for the 11-qubit problem and $\approx$~0.04 for the 8-qubit problem, which are higher compared to random sampling ($3/2^{11} \approx 0.001$ for the 11-qubit problem and $4/2^{8} \approx 0.016$ for the 8-qubit problem).

\section{Conclusion and outlook} \label{subsec:qaoa_outlook}
This work proposes and studies a procedure to obtain Pareto-optimal solutions of MCOPs in a manner amenable to near-term quantum computers via VQAs. By scalarizing the multi-objective cost function in QUBO form and variationally optimizing it with a quantum computer, the procedure allows the Pareto front of MCOPs to be recovered and visualized in an intuitive manner. Focusing on the practically relevant network routing problem, which is an instance of the multi-objective shortest-path problem, efficient QUBO formulations for practically relevant objectives were detailed. Analytical scaling analyses on the resources required for this procedure point towards implementation feasibility for real-world problem instances, for networks with bounded maximum degrees. This is supplemented by numerical simulations in small scales, which show that high-quality results can be obtained by systematically increasing the depth of the QAOA ansatz. Finally, we tested our approach on an ion trap quantum computer, obtaining consistent results that verify its applicability, constituting one of the first studies of multi-objective optimization on actual quantum computers.

Our suggested framework can be readily generalized to incorporate other MCOPs and VQAs, as long as QUBO formulations of the individual objectives are available. Firstly, to further tailor the scheme we have developed for MCOPs, more general scalarization choices beyond the simple linear scalarization approach could be explored. Examples are the Chebyshev and epsilon-constraint scalarization schemes, which are more powerful and versatile, but incur higher computational costs in the form of additional inequality constraints \cite{collette2004multiobjective, emmerich2018tutorial}. 

We can also ask whether it is possible to more directly leverage the fact that QAOA works with quantum superpositions of candidate solutions to the optimization problem. Rather than optimizing based on the energy directly, it may be fruitful to consider modifying the cost function to one that favours states which contain Pareto-optimal configurations with high probability, which shares conceptual similarities with recently developed methods such as the Conditional Value-at-Risk technique \cite{barkoutsos2020improving}. Such a modification can be performed efficiently on the sample, since the Hamiltonian corresponding to QUBO problems only contains terms that fully commute with one another.

Another improvement is in the direction of tailored state initializations and mixer designs. Due to the large fraction of infeasible solutions in the state space, the default initial state of the QAOA procedure -- an equal superposition of all computational basis states -- is highly suboptimal. Ideally one would like to be able to (a) prepare an initial state that is an equal superposition over all feasible solutions, and (b) implement a mixer Hamiltonian that does not induce transitions to infeasible solutions, so that the search is fully restricted to the feasible subspace. In the context of the network routing problem we leave this as a question for future work, and note that \cite{zhang2021qed} describes QAOA mixer Hamiltonians for network flow optimization. We also remark that the two aforementioned requirements (a) and (b) may carry a significant resource footprint, depending on the complexity of preparing the initial state and the polynomial degree of the mixer Hamiltonian.


Even on classical computers, MCOPs are generally NP-hard optimization problems that are solved with metaheuristic optimization algorithms such as particle swarm and genetic algorithms (c.f. Appendix.~\ref{appendix:moop}). Can our scheme, or VQAs in general, outperform them in some respect at large scales that are challenging for classical approaches? Due to the difficulty to theoretically analyze the performance of QAOA at large scales, and the intractability of classical simulations of quantum computations, general claims about performance likely must be made empirically, on real quantum computers. While we were able to implement and obtain solutions for the procedure for small-scale problems, applications in practical settings will require larger and more reliable quantum computers than those available today. Our work serves as a first step towards answering this question, and motivates the need for large-scale experiments on quantum computers.

\begin{acknowledgments}
We thank Adrien Vandenbroucque, Tommaso Demarie, Too Huseh Tien, and Leow Kian Siang for their valuable discussions and feedback on the manuscript, and Leonardo Disilvestro for help on the experiments on quantum computers. We further acknowledge the support of Amazon Web Services (AWS) for the computations executed using Amazon Braket. This work was supported by MINDEF and DSTA.
\end{acknowledgments}

\clearpage
\appendix


\section{Survey of relevant work} \label{appendix:survey}
This section briefly discusses relevant work in the usage of near-term quantum algorithms in solving MCOPs and the shortest path problem.

Applications of quantum annealing to the shortest path problem were first studied in \cite{krauss2020solving}, which explored different QUBO formulations of the shortest path problem. The authors of that work provide numerical and empirical evidence on its performance on a quantum annealer. With the same formalism, \cite{hauke2021dominant} studied a related problem in the context of chemistry, by again solving a relatively large problem on a quantum annealer, providing numerical evidence on its efficiency with increasing system size through a polynomially (rather than exponentially) vanishing ground state energy gap.

On the other hand, the application of near-term quantum algorithms to MCOPs is a relatively unexplored area. \cite{baran2016multiobjective} provides theoretical arguments verifying that adiabatic quantum algorithms can be applied to find Pareto-optimal points corresponding to the solution of scalarized objective functions. Most recently, \cite{urgelles2022multi} applied QAOA to the multi-objective routing problem in the context of 6G communication networks by sequentially solving single-objective problems with a lexicographic ordering method.

In relation to the above references, our work explores the solution of the multi-objective routing problem with QAOA through scalarization, and provides in-depth analysis on the approach's scalability and performance. The approach can be generalised to VQAs beyond QAOA, MCOPs beyond the routing problem, and techniques beyond linear scalarization. To the best of our knowledge, experiments on actual quantum hardware to solve multi-objective optimization problems (described in Section.~\ref{sec:qpu_results}) are new.

\section{Multi-objective Optimization} \label{appendix:moop}
This section provides a brief review of relevant notions in MOOPs and their solutions.

Given $m$ objective functions $f_i : S \rightarrow{} \mathbb{R}$, $i = 1,...,m$ that map objects from a state space $S$ to real numbers, multi-objective optimization asks for states that are optimal with respect to all of the objectives $\{ f_i \}$. Depending on their forms, the objectives generally produce competing effects with one another, so states that are simultaneously optimal in all objectives generally do not exist. In this case, we ask instead for a set of \textit{Pareto-optimal/efficient} states, which are optimal in the sense that no other states which improve on at least one individual objective without deteriorating in others can be found. The set of all Pareto-optimal states is also conventionally referred to as the $\textit{Pareto front}$, and they are always located at the boundary of the region occupied by the set of states \cite{boyd2004convex}. When the state space $S$ is finite -- as is the case for problems we consider -- this is called a \textit{multi-objective combinatorial optimization problem} (MCOP). Depending on the form of the objectives, MCOP's are generally difficult, belonging to the class of NP-hard problems \cite{back1996evolutionary,coello2007evolutionary}.

Classically, the most common approach to solve multi-objective optimization problems is with scalarization techniques. Instead of optimizing the set of objectives $\{ f_i \}$ directly, scalarization techniques aggregate them into a single objective function, thereby converting a multi-objective problem into a single-objective one. This amounts to projecting an $m$ dimensional vector onto a line, and finding the optimal state within it. Depending on the forms of $\{ f_i \}$ and the resources required to compute the objectives, different scalarization techniques are employed -- this is a well-studied subject in the literature \cite{collette2004multiobjective, emmerich2018tutorial}. For our purposes, we consider linear scalarization, where the final aggregated objective is a convex sum of the $m$ objectives:
\begin{equation} \label{eq:linear scalarization}
    f_{\text{linear}}(x) = \sum^m_{i=1} w_i f_i(x) \qquad x \in S,
\end{equation}
where the weights $w_i$ are real numbers. This corresponds to the projection of the $m$ dimensional vector onto a straight line with gradient $\vec{w}$. A property of this procedure is that a solution of the linearised problem is also a point on the Pareto front, guaranteeing the Pareto-optimality of the linearised problem's solution. Therefore, by optimizing over different choices of the weights, we can in principle build up a set of different points in the Pareto front.

While this is one of the simplest scalarization techniques, it preserves the quadratic form of the objectives and constraints in our MCOP, and also allows the scalarization weights $w_i$ to be interpreted as a priori preferences that the decision maker can select. In the case where the objectives are convex, it can be shown that every state in the Pareto front corresponds to a solution of a linearised problem. However, this simple linear scalarization may not be sufficient to capture the entire Pareto front if the objectives are concave, where some points in the Pareto front may never be captured as solutions of any linearised problem. This can be overcome with more advanced scalarization techniques such as Chebychev scalarization \cite{collette2004multiobjective, emmerich2018tutorial}, at the expense of introducing inequality constraints. In the context of our work, if a non-linear scalarization scheme is used, the resulting cost Hamiltonian may contain higher-order terms, resulting in cost unitary circuit with greater depth due to the need to implement higher-order gates. Alternatively, one can consider a more general variational ansatz, and optimize over a non-linear cost Hamiltonian.

Finally, to optimize the scalarized objective classically, suitable heuristic optimization algorithms are usually employed. Algorithms that work very well in practice are population (e.g. ant colony methods) and evolutionary or genetic-based algorithms, such as NSGA-II \cite{collette2004multiobjective, emmerich2018tutorial}.

More broadly, approaches to solve MOOPs can be broadly categorized into three classes:
\begin{itemize}
    \item \textit{A priori methods} take the preferences (e.g. the weights/relative importance of each objective) of the engineer/decision maker into account prior to the optimization, and adapts the optimization process based on this preference. Scalarization falls into this class of methods, where the weights of the objectives are specified in an a priori manner.
    
    \item \textit{A posteriori methods} aim to solve for either a representative subset or all possible Pareto-optimal solutions, only taking preferences of the decision maker into account after the optimization process.
    
    \item \textit{Interactive methods} are adaptive and iterative methods that require the continuous interaction of the decision maker at each step of the optimization process.
\end{itemize}

As discussed in the main text, our scheme belongs to the class of a priori or a posteriori methods, depending on whether the choice of scalarization weights is made explicitly.

\section{QUBO Formulation of the multi-objective shortest path problem} \label{appendix:shortest}
Following the problem statement in Section~\ref{sec:network_routing}, this appendix details the QUBO formulation of the multi-objective shortest path problem. We begin by providing an encoding of the state space of the problem in terms of binary variables, before detailing the form of the quadratic cost function to be optimized. More detailed analyses in the context of annealing, including an extension to directed graphs, can be found in \cite{krauss2020solving}.

An encoding of the problem can be achieved with $|V| + |E| - 2$ binary variables representing each state $x$ in the state space \cite{krauss2020solving} (We will use $x_i$'s and $x_{ij}$'s interchangeably with $x$ to denote states when the context is clear from now on.). The first $|V|-2$ variables $x_1, x_2, ..., x_{|V|-2}$ correspond to the nodes in the graph excluding the source and destination nodes, while the remaining $|E|$ variables $x_{ij}$, where $(i,j) \in E$, correspond to the edges. With this encoding, we consider objectives that take the following forms:
\begin{itemize}
    \item \textbf{Node cost: } Associates each node in the graph with a cost, depending on the node's weight $V_i$.
    \begin{equation} \label{eq:node_cost}
        E_{\text{node}}(x) = \sum_{i \in V} V_{i} x_{i}
    \end{equation}
    
    \item \textbf{Edge cost: } Associates each edge in the graph with a cost, depending on the edge's weight $E_{ij}$. 
    \begin{equation} \label{eq:edge_cost}
        E_{\text{edge}}(x) = \sum_{(i,j) \in E} E_{ij} x_{ij}
    \end{equation}
\end{itemize}
With only one objective (for instance, the minimization of the distance between two points of a graph, which is an edge cost), the problem reduces to a single-objective shortest path problem which can be solved efficiently in polynomial time by Dijkstra's algorithm \cite{dijkstra1959}, which takes $O(|E|+|V|\log|V|)$ steps. The presence of multiple objectives constitute a MCOP.

Eqs.~(\ref{eq:node_cost}) and (\ref{eq:edge_cost}) allow us to compute the cost vector associated with a state $x$. However, of the $2^{|V| + |E|-2}$ possible states, not all of them represent actual paths. A valid path is one that (1) starts from a specified source node $s$, (2) ends at a specified destination node $d$, and (3) has no broken links or branches along the path from $s$ to $d$. States that satisfy these criteria are called \emph{feasible}, and \emph{infeasible} otherwise. These 3 constraints can be enforced as quadratic penalty terms added to our cost function, so that infeasible solutions have higher total energies than feasible ones:

\begin{itemize}
    \item \textbf{Source constraint: } Penalizes paths that do not have exactly one edge connected to the source node:
    
    \begin{equation} \label{source_constraint}
        E_s(x) = -x^2_s + (x_s - \sum_j x_{sj})^2,
    \end{equation}
    where $s$ is the index of the source node, and the sum is over all edges that are connected to node $s$.  This term has a minimum value of $-1$, which occurs for states where the source node is used ($x_s = 1$), and there is only one way of leaving the source node (the bracketed term is equal to zero).
    
    \item \textbf{Destination constraint: } Penalizes paths that do not have exactly one edge connected to the destination node:
    
    \begin{equation} \label{destination_constraint}
        E_d(x) = -x^2_d + (x_d - \sum_j x_{dj})^2,
    \end{equation}
    where $d$ is the index of the destination node, and the sum is over all edges that are connected to node $d$. This term has a minimum value of $-1$, similar to the source constraint.
    
    \item \textbf{Path constraint: } Penalizes paths that do not have exactly 2 edges connected to intermediate nodes: $E_{\text{path}} = \sum_{i \in V} E_i$, such that for each intermediate node $i$:
    
    \begin{equation} \label{path_constraint}
        E_i(x) = (2x_i - \sum_j x_{ij})^2,
    \end{equation}
    where the sum is over all edges that are connected to node $i$. This has a minimum value of 0, which occurs for states where for all intermediate nodes used (for which $x_i = 1$), the node degree is equal to 2.
\end{itemize}
A solution of the multi-objective shortest path problem is then a path $x$ that achieves the minimum of the penalties Eqs.~(\ref{source_constraint}),(\ref{destination_constraint}), and (\ref{path_constraint}) and is Pareto-optimal with respect to the node and edge costs.

As mentioned in the main text, the connectivity of the graph associated with the sum of the path constraints is precisely that of the problem's middle graph, which we review and exploit in Appendices \ref{appendix:middle_graph_properties} and \ref{appendix:proof} respectively. Furthermore, the constraints Eqs.~(\ref{source_constraint}) and (\ref{destination_constraint}) only incur a small number of linear and quadratic terms (dependent on the local connectivity of the source and destination nodes) which we safely ignore for conciseness.

\section{Details on Graph Theory, the Middle Graph, and the Proof of Eq.~(\ref{eq:edgecolor_bound})} \label{appendix:graph_theory}
Here, we list down notions in graph theory used in the resource estimation part of the main text, and provide a proof for Eq.~(\ref{eq:edgecolor_bound}). 

As mentioned in Section.~\ref{sec:network_routing}, the correspondence $G_H = M(G)$ implies that the solution of a shortest path problem defined on a network $G$ can be mapped to the ground state of a Hamiltonian/Ising model $H$ with graph $G_H = M(G)$. Ignoring the quadratic terms resulting from the source and destination nodes, which only contributes a negligible constant overhead, Eq.~(\ref{eq:edgecolor_bound}) implies that the implementation of the cost unitary is efficient, depending only linearly on the maximum degree of the network $G$. 

\subsection{Notions in graph theory}
For an undirected graph $G = (V(G), E(G))$, the degree of a vertex $u \in V(G)$, denoted $\text{deg}(u)$, is the number of edges that are incident to the vertex. The maximum degree, denoted $\Delta_G$, is the maximum of its vertices' degrees $\Delta_G = \max_{u \in V(G)} \text{deg}(u)$. The degree sum formula relates them to the number of edges of $G$:
\begin{equation}
    |E(G)| = \frac{1}{2}\sum_{u \in V(G)} \text{deg}(u) \leq \frac{1}{2}|V(G)|\Delta_G.
\end{equation}

The line graph of $G = (V(G), E(G))$ is the graph $L(G) = (E(G), E')$ with the edges of $G$ as vertices, such that they are adjacent if and only if their corresponding edges are incident on the same vertex. The maximum degrees of $G$ and $L(G)$ are related by:
    \begin{equation} \label{eq:graph_lg_deg}
        \Delta_{L(G)} \leq 2 \Delta_G - 2.
    \end{equation}

The middle graph of $G = (V(G), E(G))$ \cite{hamada1976traversability} is the graph $M(G) = (V(G) \cup E(G), E')$, with vertices $u,v$ that are adjacent if either:
\begin{enumerate}[label=(\alph*),nolistsep]
    \item $u$ is a vertex in $G$ and $v$ is an edge in $G$ incident to $u$, or
    \item $u$ and $v$ are edges in $G$ that both incident on the same vertex.
\end{enumerate}

The endline graph of $G$, denoted $G^+$, is defined as the graph obtained from $G$ by adding to each of its nodes an end-vertex (i.e. an edge that is connected to a single node). The attachment of an end-vertex to each node immediately implies that:
\begin{equation} \label{eq:graph_endline_deg}
    \Delta_{G^{+}} = \Delta_G + 1.
\end{equation}

An edge coloring of a graph $G$ is an assignment of colors to its edges, such that no two edges sharing a vertex can have the same color. The edge chromatic number/chromatic index of $G$, denoted $\chi'(G)$, is the minimum number of such colors needed. Vizing's theorem relates the edge chromatic number of $G$ with the maximum degree for any graph:
\begin{equation}
    \Delta_G \leq \chi'(G) \leq \Delta_G + 1.
\end{equation}

\subsection{Properties of the middle graph} \label{appendix:middle_graph_properties}
We introduce some properties on $M(G)$ that will be used:
\begin{enumerate}
    \item The number of vertices and edges of $M(G)$ are respectively:
        \begin{equation} \label{eq:mg_nodes}
            |V(M(G))| = |V(G)| + |E(G)| \leq \frac{1}{2}|V(G)|(2 + \Delta_G),
        \end{equation}
        \begin{equation} \label{eq:mg_edges}
            |E(M(G))| = \frac{1}{2}\sum_{u \in V(G)} \text{deg}(u)^2 \leq \frac{1}{2}|V(G)| \Delta_G^2.     
        \end{equation}

    \item The middle graph and the endline graph of the line graph of $G$ are isomorphic \cite{hamada1976traversability}:
        \begin{equation}
            M(G) \cong L(G)^{+}.
        \end{equation}
    This immediately implies that their maximum degree are also equal:
    \begin{equation}
        \Delta_{M(G)} = \Delta_{L(G)^{+}}
    \end{equation}

\end{enumerate}

\subsection{Proof of $\chi'(M(G)) \leq 2 \Delta_G$} \label{appendix:proof}
The following chain of (in)equalities are true:
\begin{align}
\chi'(M(G)) & \leq \Delta_{M(G)} + 1 \\
& = \Delta_{L(G)^{+}} + 1 \\
& = \Delta_{L(G)} + 2 \\
& \leq 2 \Delta_G,
\end{align}
where applying Vizing's theorem yields the first inequality, $M(G) \cong L(G)^{+}$ yields the second equality, Eq.~(\ref{eq:graph_endline_deg}) yields the third inequality, and Eq.~(\ref{eq:graph_lg_deg}) yields the final equality. \qed

\section{QUBO Formulation of the network routing problem} \label{appendix:network}
Following Section.~\ref{sec:network_routing} and Appendix.~\ref{appendix:shortest}, this section details how the objectives of the multi-objective network routing problem (path loss, node delay, data rate, and bit-error) are formulated in our work, in particular how each objective can be modelled as either a node, Eq.~(\ref{eq:node_cost}), or edge, Eq.~(\ref{eq:edge_cost}), cost to be minimized. Network parameters used to generate problem instances considered in numerical simulations and experiments on quantum computers are summarized in Table.~(\ref{tab:params}).

\begin{table}[ht]
	\begin{center}
		\begin{tabular}{|c|c|}
			\hline
			Parameters & Values \\
			\hline
			Network coverage area (Fully-connected) &  $1000 \text{m} \times 1000 \text{m}$ \\
   Network coverage area (Square lattice) &  $1000 \text{m} \times 2000 \text{m}$ \\
            Node delay, $\Delta$ & $1 ~ \text{ms}$ \\
            Path loss exponent, $\alpha$ & 2.7 \\
            Transmitting power, $P_T$ & $50 ~ \text{W}$  \\
            Carrier wavelength, $\lambda_c$ & $1.2 ~ \text{m}$\\
            Carrier frequency, $f_c$ & $250 ~ \text{MHz}$ \\
            Modulation & QPSK \\
            BER Gaussian noise mean, $\mu$ & $-90 ~ \text{dBm}$ \\
            BER Gaussian noise std. dev., $\sigma$ & $10 ~ \text{dBm}$ \\
 		\hline
		\end{tabular}
		\caption{\label{tab:params} Network parameters}
	\end{center}
\end{table}

\subsection{Node Delay}
The node delay quantifies the total time delay incurred by intermediate processing steps at relay stations along a signal's trajectory from sender to receiver, and can be written as a linear objective function:
\begin{equation}
	E_{\Delta}(x) = \sum_{i\in V} \Delta_i x_i,
\end{equation}
where $\Delta_i$ is the time delay incurred by node $i$. In our numerical simulations, we further assume that every node in the network incurs the same node delay of $\Delta_i = \Delta = 1 \text{ms}$, so that the node delay is proportional to the number of nodes traversed by the signal.

\subsection{Path Loss}
The Path Loss metric accounts for the loss in the amplitude of the electromagnetic signal during transmission between stations, which increases with distance travelled \cite{steele1999mobile,Fernandez2014pathloss}. Between nodes $i$ and $j$, it is given by:
\begin{equation}
    L_{ij} = \frac{P_T}{P_R} = \left(\frac{4\pi d_{ij}}{\lambda_c}\right)^{\alpha},
\end{equation}
where $P_T = 50 \text{W}$ is the power at the transmitting station, $P_R$ the power at the receiving station, $\alpha = 2.7$ the path loss exponent (typically between 2 in free space and 4 in a lossy environment), $d_{ij}$ the distance between nodes $i$ and $j$, and $\lambda_c = 1.2\text{m}$ the carrier wavelength (resulting in the carrier frequency of $f_c = 250 \text{MHz}$, in the VHF regime). The resulting total path loss can be written as the sum of linear terms of the form:
\begin{equation}
	E_{\text{L}}(x) = \sum_{(i,j)\in E}L_{ij}x_{ij}
\end{equation}

\subsection{Symmetrized Bit Error Rate}
The Bit Error Rate accounts for the probability of introducing an error into the message during the signal processing phase at a node. This error takes the shape of bit flip in the message and depends on the power of the signal and the noise at the receiving node:
\begin{align}
	p_{ij} = \frac{1}{2}\left(1-\sqrt{\frac{R_{ij}}{R_{ij}+1}}\right),
\end{align}
where $R_{ij} = \frac{P_{R,ij}}{\log_2(M)P_{N,ij}}$, $M=4$ for QPSK modulation, and $P_{R,ij} = P_T \left(\frac{\lambda_c}{4\pi d_{ij}}\right)^{\alpha}$. We model the noise at each node $P_{N,i}$ to follow a normal distribution with mean $\mu = -90\text{dBm}$ and standard deviation $\sigma = 10\text{dBm}$, and additionally average (and thus symmetrize) the noise at neighbouring nodes, i.e. $P_{N,ij} = \frac{1}{2}(P_{N,i}+P_{N,j})$. This yields a linear cost function associated with the connecting edge of the form:
\begin{equation}
	E_{\text{BER}} = \sum_{(i,j)\in E}p_{ij}x_{ij},
\end{equation}
so that the total error probability is the sum of the individual error probabilities for each selected edge. It is a first order approximation of the exact total error (equivalent to assuming errors do not occur twice on the same bit).

\subsection{Data Rate}
The Data Rate metric quantifies the maximum data transfer rate that can be achieved across a transmission path. The Nyquist data rate for noiseless signals can be described by $\Gamma_{ij} = 2B_{ij}\text{log}_2(M)$, with $M = 4$ the modulation level for QPSK and $B_{ij}$ the bandwidth of the signal transmitting between nodes $i$ and $j$. We model the total available data rate between two nodes to be the base rate of $5 \text{Mbit/s}$ minus the utilised data rate, such that it results in total available data rates that are uniformly distributed between $0$ and $5 \text{Mbit/s}$, in steps of $50 \text{kbit/s}$.

The link with the minimum data rate determines the overall data rate along a transmission path, forming a bottleneck. In other words, we wish to penalize transmission paths with low minimum data rates. To express this metric in QUBO form as a linear sum of edge costs, we use a negative exponential function to weight the data rates $\Gamma_{ij}$: 
\begin{equation}
	E_{\Gamma} = \sum_{(i,j)\in E}x_{ij}\cdot e^{-\beta\Gamma_{ij}},
\end{equation}
so that sub-optimal links with low data rates are penalized heavily. Here, $\beta$ is a graph-dependent coefficient that should be chosen such that for any choice of transmission path $x$, the dominant contribution to $E_{\Gamma}$ corresponds to the minimum data rate along the path. It can be estimated by either considering the link in $G$ with the smallest possible data rate, or by determining the value of $c(\beta) = \sum_{\mathrm{x \in \text{Paths}}}\left(e^{-\beta R^x_{\text{min}}} - \sum_{j \neq \text{min}} e^{-\beta R^x_j}\right)$ by sampling over different paths. $R^x_\text{min}$ is the minimum data rate of path $x$, and $R^x_j$ are the data rates on all other edges in $x$.

\section{Specifications of problem instances in QPU computations} \label{appendix:problem_instances} 
This section specifies the parameters used to generate problem instances solved in Section.~\ref{sec:qpu_results} with the IonQ quantum computer, with the four objectives defined in Appendix.~\ref{appendix:network}. Nodes of the network are distributed within a $(1000 \text{m} \times 1000 \text{m})$ area for the fully-connected networks and $(1000 \text{m} \times 2000 \text{m})$ for the square lattice networks.

\subsection{Fully-Connected Network}
The fully connected network is a 4-node graph with 5 feasible solutions, of which 4 are Pareto-optimal. 8 qubits are required to encode this problem.

\begin{table}[ht]
	\begin{center}
		\begin{tabular}{|c|c|c|c|}
			\hline
			node $i$ & coordinates & noise level & node delay \\
			&[m]& $P_{N,i}$ [dBm] & $\Delta$ [ms]\\
			\hline
			0 & (-181, 88) & -93.665 & 1.0 \\
			1 & (-82, 674) & -94.609 & 1.0 \\
			2 & (589, 84) & -81.811 & 1.0 \\
			3 & (645, 664) & -85.622 & 1.0 \\
			\hline
		\end{tabular}
		\caption{\label{tab:fund_node_full_b} Node properties and parameters of the fully connected network.}
	\end{center}
\end{table}

\begin{table}[ht]
	\begin{center}
		\begin{tabular}{|c|c|c|c|c|c|}
			\hline
			edge & distance & data rate & path loss  & $R_{ij}$ & bit error \\
			$(i,j)$ &$d_{ij}$ [m]& $\Gamma_{ij}$ [kbit/s] & $L_{ij}$ [dB] &[dB]&$p_{ij}$\\
			\hline
			(0, 1) & 594.757 & 4250 & 102.448 & 35.643 & 6.82e-05 \\
			(0, 2) & 770.565 & 2350 & 105.485 & 23.042 & 1.24e-03 \\
			(0, 3) & 1007.517 & 1801 & 108.629 & 23.351 & 1.15e-03 \\
			(1, 2) & 893.738 & 3850 & 107.223 & 21.355 & 1.82e-03 \\
			(1, 3) & 727.298 & 1350 & 104.807 & 27.289 & 4.66e-04 \\
			(2, 3) & 582.504 & 1900 & 102.204 & 25.087 & 7.73e-04 \\
			\hline
		\end{tabular}
		\caption{\label{tab:edge_metrics_full_b} Edge properties and parameters of the fully connected network.}
	\end{center}
\end{table}

\subsection{Square Lattice Network}
The square lattice network is a 6-node graph with 4 feasible solutions, of which 3 are Pareto-optimal. 11 qubits are required to encode this problem.

\begin{table}[ht]
	\begin{center}
		\begin{tabular}{|c|c|c|c|}
			\hline
			node $i$ & coordinates & noise level & node delay \\
			&[m]& $P_{N,i}$ [dBm] & $\Delta$ [ms]\\
			\hline
			0 & (157, -93) & -85.058 & 1.0 \\
			1 & (73, 316) & -92.388 & 1.0 \\
			2 & (425, -102) & -91.186 & 1.0 \\
			3 & (408, 469) & -93.737 & 1.0 \\
			4 & (980, -57) & -84.012 & 1.0 \\
			5 & (999, 635) & -91.733 & 1.0 \\
			\hline
		\end{tabular}
		\caption{\label{tab:fund_node_sq_c} Node properties and parameters of the square lattice network.}
	\end{center}
\end{table}

\begin{table}[ht]
	\begin{center}
		\begin{tabular}{|c|c|c|c|c|c|}
			\hline
			edge & distance & data rate & path loss  & $R_{ij}$ & bit error\\
			$(i,j)$ &$d_{ij}$ [m]& $\Gamma_{ij}$ [kbit/s] & $L_{ij}$ [dB] & $p_{ij}$ [dB]&\\
			\hline
			(0, 1) & 417.987 & 300 & 98.312 & 32.999 & 1.25e-04 \\
			(0, 2) & 268.148 & 3950 & 93.107 & 37.993 & 3.97e-05 \\
			(1, 3) & 368.637 & 3600 & 96.839 & 40.151 & 2.41e-05 \\
			(2, 3) & 571.589 & 2200 & 101.982 & 34.274 & 9.34e-05 \\
			(2, 4) & 556.17 & 3250 & 101.661 & 28.579 & 3.46e-04 \\
			(3, 5) & 614.043 & 650 & 102.822 & 33.777 & 1.05e-04 \\
			(4, 5) & 693.11 & 3000 & 104.242 & 26.081 & 6.15e-04 \\
 			\hline
		\end{tabular}
		\caption{\label{tab:edge_metrics_sq_c} Edge properties and parameters of the square lattice network.}
	\end{center}
\end{table}

\section{Additional numerical results on parameter initialization} \label{appendix:initializations}
\begin{figure}[ht]
    \centering
    \includegraphics[width=1.1\columnwidth]{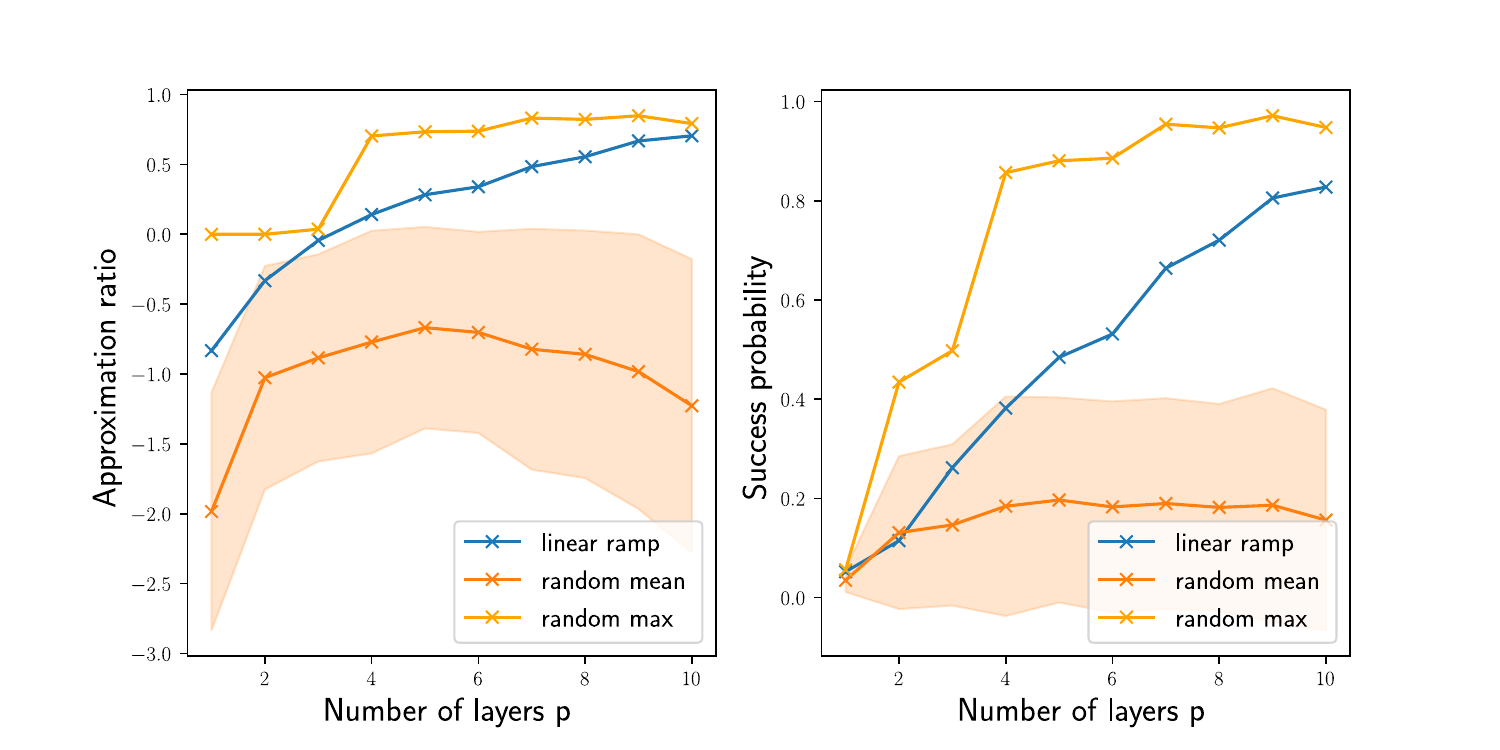}
    \caption{Scaling of the approximation ratio and success probability with the number of layers for different initializations, for one specific problem. To produce the random initializations curves, we perform $100p$ random initializations and optimize with QAOA. The lighter region displays standard deviation around the mean. The average result from $100p$ random initializations is labelled as `random mean', the best result from the $100p$ random initializations is labelled as `random max' (to refer to the maximum approximation ratio achieved), and the (one) result from the linear ramp initialization is labelled as `linear ramp'.}
    \label{fig:scaling_init}
\end{figure}

In this section, we briefly verify the validity of the linear ramp parameter initialization scheme used in our numerical simulations, which is a heuristic choice that linearly ramps up $\gamma$'s and ramps down $\beta$'s, based on the analogy between QAOA and quantum annealing \cite{zhou2020quantum}. Fig.~(\ref{fig:scaling_init}) displays numerical scaling results comparing this initialization (labelled as `linear ramp') with random initializations for different values of $p$. The number of random initializations are scaled as $100p$, which in our numerical examples is sufficient to capture the global minima in most cases. The best result from the $100p$ random initializations (labelled as `random max' to refer to the maximum approximation ratio achieved) can therefore be taken as the global optimum of the problem. For illustration, the problem instance is chosen to be a 5-node triangular lattice with 1 column and 3 rows, an 8-qubit problem.

We observe that the linear ramp initialization indeed greatly outperforms random sampling on average -- both the approximation ratio and success probabilities approach the global optima in a monotonic fashion as $p$ increases, compared to an average random initialization run, which saturates and even deteriorates for large $p$. We will therefore proceed to conduct all experiments presented in this work with the linear ramp initialization, unless stated otherwise.

\section{Specification of problem instance of Fig.~(\ref{fig:pareto_qaoa})} \label{appendix:fig_problem}
Table.~\ref{tab:plot_metrics} provides the complete specification of the triangular lattice graph and the multi-objective cost function for the problem instance used in Fig.~(\ref{fig:pareto_qaoa}). That is, each of the 4 objective functions takes the form:
\begin{equation} \label{eq:fig_obj}
    E^C_i(x) = \sum_{i} h_i x_i + \sum_{(i,j)} J_{ij} x_i x_j,
\end{equation}
where $h_i$ corresponds to node weights and $J_{ij}$ corresponds to edge weights with values as assigned in the table.
\begin{table}[ht]
	\begin{center}
		\begin{tabular}{|c|c|c|c|c|}
			\hline
			Node (i) & Obj. 1 & Obj. 2 & Obj. 3  & Obj. 4  \\
			\hline
			(0) & 1.5 & 0 & 0.5 & 0  \\
			(1) & 0.5 & 0 & 1   & 0  \\
			(2) & 1.5 & 0 & 1.5 & 0  \\
			(3) & 0.2 & 0 & 2   & 0  \\
			(4) & 1   & 0 & 1   & 0  \\
			(5) & 0.5 & 0 & 1.5 & 0  \\
                \hline
                Edge (i,j) & Obj. 1 & Obj. 2 & Obj. 3  & Obj. 4  \\
                \hline
   			(0,1) & 0 & 0.5 & 0 & 2.5  \\
			(0,3) & 0 & 1.5 & 0 & 0.1  \\
			(1,2) & 0 & 1.6 & 0 & 0.5  \\
			(1,4) & 0 & 1   & 0 & 1.5  \\
			(1,3) & 0 & 3   & 0 & 0.5  \\
			(2,5) & 0 & 1.6 & 0 & 0.6  \\
                (2,4) & 0 & 2   & 0 & 0.5  \\
			(3,4) & 0 & 1   & 0 & 0.5  \\
			(4,5) & 0 & 0.6 & 0 & 1.6  \\
			\hline
		\end{tabular}
		\caption{\label{tab:plot_metrics} Weights of cost terms of the triangular lattice problem instance of Fig.~(\ref{fig:pareto_qaoa}) for all 4 objectives, labelled according to their corresponding edges or nodes. 6 node weights and 9 edge weights are present. The problem can then be specified via Eq.~(\ref{eq:fig_obj}).}
	\end{center}
\end{table}

\section{Fraction of feasible vs. infeasible configurations}\label{appendix:Feasible_Infeasible}
Here we provide an intuitive explanation for the observation made in Section.~\ref{subsec:Pareto_plots} that the ratio of the number of feasible solutions to the number of infeasible solutions is a decreasing function of the network size. We will illustrate this from the standpoint of the constraint described by Eq.~(\ref{path_constraint}), which enforces that any intermediate vertex used in a path through the network must have degree 2. All feasible solutions must, as a necessary condition, satisfy this particular constraint. Our argument will be based on how we can construct infeasible configurations that violate this constraint, starting from feasible ones. 

Suppose we have a feasible configuration that uses intermediate vertex $j$ (implying $x_j=1$), with the path entering vertex $j$ through the edge $(i, j)$, and leaving through the edge $(j, k)$ (implying that $x_{ij} = x_{jk} = 1$). There are multiple infeasible configurations in the state space that can be obtained from this feasible configuration by simply flipping the value of one variable. For instance, if we set $x_j=0$ in the example above, and leave all other variables unchanged, we have an infeasible configuration that describes a path using the edges $(i,j)$ and $(j,k)$, but which never enters the vertex $j$. Similarly, if we instead set $x_{ij}=0$, and leave everything else unchanged, we would have a path that leaves vertex $j$, but never enters it. If there are more than just two possible paths in or out of vertex $j$, we could also have solutions where the degree of the vertex is larger than two, which would also represent an infeasible configuration.

The examples just discussed would violate the constraint of Eq.~(\ref{path_constraint}), and hence they represent infeasible configurations. As we see, for every feasible configuration, we can minimally `perturb' it by flipping the value of individual variables to obtain infeasible configurations. However, we could build `even worse' infeasible configurations by introducing such perturbations in multiple locations simultaneously, not only around a single vertex or edge. The number of combinations of infeasible configurations we can generate in this way grows exponentially in the number of variables (i.e. in the number of vertices and edges). We therefore conclude that the fraction of feasible to infeasible configurations shrinks exponentially in the problem size.

\newpage
\bibliographystyle{apsrev4-2}
\bibliography{reference}

\end{document}